\documentclass[sigconf,ACM]{acmart}

\AtBeginDocument{%
  \providecommand\BibTeX{{%
    \normalfont B\kern-0.5em{\scshape i\kern-0.25em b}\kern-0.8em\TeX}}}

\usepackage{enumitem}
\usepackage{multirow}
\usepackage{subfigure}
\usepackage[ruled,vlined,linesnumbered]{algorithm2e}

\usepackage{xcolor}

\copyrightyear{2023} 
\acmYear{2023} 
\setcopyright{acmlicensed}\acmConference[WWW '23]{Proceedings of the ACM Web Conference 2023}{April 30-May 4, 2023}{Austin, TX, USA}
\acmBooktitle{Proceedings of the ACM Web Conference 2023 (WWW '23), April 30-May 4, 2023, Austin, TX, USA}
\acmPrice{15.00}
\acmDOI{10.1145/3543507.3583422}
\acmISBN{978-1-4503-9416-1/23/04}

\begin{document}

\title{Cooperative Retriever and Ranker in Deep Recommenders}
\author{Xu Huang$^1$, Defu Lian$^{1,2,3}$, Jin Chen$^4$, Zheng Liu$^{5}$, Xing Xie$^{5}$, Enhong Chen${^{1,2,3}}$}
\authornote{Defu Lian is the Corresponding Author}

\affiliation{%
    \institution{$^1$School of Computer Science, University of Science and Technology of China \city{Hefei} \country{China}}
    \institution{$^2$School of Data Science, University of Science and Technology of China \city{Hefei} \country{China}}
    \institution{$^3$State Key Laboratory of Cognitive Intelligence \city{Hefei} \country{China}}
    \institution{$^4$University of Electronic Science and Technology of China \city{Chengdu} \country{China}}
    \institution{$^5$Microsoft Research Asia \city{Beijing} \country{China}}
}
\email{xuhuangcs@mail.ustc.edu.cn, {liandefu,cheneh}@ustc.edu.cn, chenjin@std.uestc.edu.cn, {zhengliu,xingxie}@microsoft.com}
% \affiliation{%
%     \city{University of Science and Technology of China}
%     % \city{Hefei}
%     \country{China}
% }

% \author{Defu Lian}
% \authornote{Corresponding author}
% \email{liandefu@ustc.edu.cn}
% \affiliation{%
%     \city{University of Science and Technology of China}
%     % \city{Hefei}
%     \country{China}
% }

% \author{Jin Chen}
% \email{chenjin@std.uestc.edu.cn}
% \affiliation{%
%     \city{University of Electronic Science and Technology of China}
%     % \city{Chengdu}
%     \country{China}
% }

% \author{Zheng Liu}
% \email{zhengliu@microsoft.com}
% \affiliation{%
%     \city{Microsoft Research Asia}
%     % \city{Beijing}
%     \country{China}
% }

% \author{Xing Xie}
% \email{xingxie@microsoft.com}
% \affiliation{%
%     \city{Microsoft Research Asia}
%     % \city{Beijing}
%     \country{China}
% }

% \author{Enhong Chen}
% \email{cheneh@ustc.edu.cn}
% \affiliation{%
%     \city{University of Science and Technology of China}
%     % \city{Hefei}
%     \country{China}
% }

\begin{abstract}
% Deep recommender systems (DRS) are intensively applied in modern web services, such as e-commerce and personalized feeds. 
% To deal with massive-scale of web content, DRS jointly leverages the retrieval and ranking operations to generate the recommendation result.
% The retriever targets selecting a small set of relevant candidates from the entire items with high running efficiency; while the ranker, usually more precise but time-consuming than the retriever, is supposed to identify the best items out of the retrieved candidates with high precision. 
% Despite the collaborative nature of both components, the retriever and ranker are usually trained in independent or poorly-cooperative ways, leading to severely limited recommendation performances when working as an entirety.

{Deep recommender systems (DRS) are intensively applied in modern web services. To deal with the massive web contents, DRS employs a two-stage workflow: retrieval and ranking, to generate its recommendation results. The retriever aims to select a small set of relevant candidates from the entire items with high efficiency; while the ranker, usually more precise but time-consuming, is supposed to further refine the best items from the retrieved candidates. Traditionally, the two components are trained either independently or within a simple cascading pipeline, which is prone to poor collaboration effect. Though some latest works suggested to train retriever and ranker jointly, there still exist many severe limitations: item distribution shift between training and inference, false negative, and misalignment of ranking order. As such, it remains to explore effective collaborations between retriever and ranker.}

{In this work, we present a novel framework for the joint training of retriever and ranker, named \textbf{CoRR} (\textbf{Co}operative \textbf{R}etriever and \textbf{R}anker). With CoRR, the retriever is improved by deriving high-quality training signals from the ranker, while the ranker is improved by learning to discriminate hard negatives sampled by the retriever. We introduce two critical techniques. Firstly, we develop an adaptive and scalable sampler based on the retriever, to generate hard negative samples for the ranker's training. Compared with the widely-used exact top-k sampling, our method effectively alleviates the issues of false negative and item distribution shift, and thus improves the ranker's discriminability. Secondly, we propose a novel asymptotic-unbiased estimation of KL divergence, which serves as the objective for knowledge distillation. The new objective can be efficiently optimized with commonly-used optimizers. More importantly, it leads to better alignment of ranking order between retriever and ranker, which helps to improve the retrieval quality. We conduct comprehensive experiments over four large-scale datasets, where CoRR outperforms both conventional DRS and the existing joint training methods with notable advantages. Our code will be open-sourced to facilitate future research.} 
\end{abstract}



\maketitle

\section{Introduction}
Recommender system plays an important role in modern web services, like e-commerce and online advertising, as it largely mitigates the information overload problem by suggesting users with personalized items according to their own interests. Thanks to the remarkable progress of deep learning, deep recommender systems (DRS) become increasingly popular in practice \cite{covington2016deep,hidasi2015session,ying2018graph, feng2022recommender,lian2020lightrec,jin2020sampling,wu2021linear,feng2023reinforcement}. Given the magnificent scale of online items, deep recommender systems call for a two-stage workflow: retrieval and ranking~\cite{covington2016deep}. Particularly, the retriever targets on selecting a small set of candidate items under a certain context (e.g., user profile and historical interactions) from the whole items with high efficiency. Typically, the retriever model learns to represent the context and items as dense embeddings, such that user's preference towards the items can be efficiently estimated based on embedding similarities, like inner product or cosine similarity. In contrast, the ranker is used to refine the most preferred items from the retrieval results. For the sake of best precision, it usually leverages highly expressive yet time-consuming networks, especially those establishing deep interactions between the context and the item (e.g., DIN~\cite{zhou2018deep}, DIEN~\cite{zhou2019deep} and DeepFM~\cite{guo2017deepfm}). 

% , for the fine-grained prediction of context-aware preference towards the retrieved items. 

%% independent, poorly cooperative: both need experiment?
%% discussion of ir-gan?
%% why top-K is bad, limited sample due to the implementation?
\subsection{Existing Problems} 
Despite the collaborative nature of the retriever and ranker, the typical training workflow of the two models lacks effective cooperation, which severely harms the overall recommendation quality. In many cases, the two models are independently trained and directly applied to the recommender systems~\cite{zhu2018learning,cen2020controllable,zhou2018deep,zhou2019deep,lian2020geography}. 
At other times, the ranker can be successively trained based on retrieval results~\cite{gallagher2019joint,hron2021component}; whereas, the retriever remains independently trained~\cite{covington2016deep}. Such training workflows are inferior due to the following reasons. 

$\bullet$ {The independent training of the retriever only leverages the historical user-item interactions, which can be limited in reality. As a result, it may suffer from the sparsity of training data, which severely restricts the retrieval quality. For another thing, the retriever is likely to generate candidate items that are not favored by the ranker; thus, it may harm the downstream performance as well.} 

$\bullet$ {The independently-trained ranker is learned with randomly or heuristically collected training samples. Such training samples can be too easy to be distinguished, making the ranker converge to a limited discriminative capacity. Besides, the item distribution will also be highly differentiated between the training and inference stages; as a result, the ranker may not effectively recognize the high-quality candidates generated by the retriever.} 

{Recent studies on multi-stage ranking models are closely related to the problem of retriever-ranker collaboration; e.g., in \cite{qin2022rankflow}, a two-pass training workflow is proposed. However, the two-pass workflow is still limited from several critical perspectives. In the forward pass, the rankers are trained by the retrieval results at the exact top-k cutoff, which is prone to false-negatives. Besides, when the retrieval cutoffs are changed during the inference stage, the rankers will face highly shifted item distributions from the training stage, which may severely harm their prediction accuracy. In the backward pass, the retrievers are trained to preserve the consistency of absolute ranking scores and to distinguish ranking results from retrieval ones; while for the sake of high-quality retrieval, it is the consistency of relative ranking order that really matters. In all, it remains a challenging problem to explore more effective collaboration mechanisms between the retriever and ranker.} 

% Recent study begins to realize the above problems and proposed to train retriever and ranker jointly \cite{qin2022rankflow} within a bi-pass workflow. However, many critical issues remain unexplored for the effective collaboration between the two modules. 

\subsection{Our Solution} 
In this work, we propose a novel framework for the cooperative training of the retriever and ranker, a.k.a. \textbf{CoRR} (\textbf{Co}operative \textbf{R}etriever and \textbf{R}anker). In our framework, the retriever and ranker are simultaneously trained within a unified workflow, where both models can be mutually reinforced. 

{$\bullet$ \textbf{Training of retriever}. On one hand, the retriever is learned from both user-item interactions via sampled softmax and the ranker's predictions via knowledge distillation~\cite{hinton2015distilling}. In a specific context, a few items are sampled firstly. %in the first place. 
Then, the ranker is required to predict the fine-grained preferences towards the sampled items. Rather than preserving the absolute preference scores, the retriever is required to generate the same ranking order for the sampled items as the ranker. To realize this goal, the KL-divergence is minimized for the softmax-normalized predictions between the retriever and ranker~\cite{hinton2015distilling}. In this case, user's preferred items, whether interacted or not, will probably get highly-rated by the ranker, while real non-interested items get lower-rated (thanks to the highly-precise ranker). As a result, such items will \textbf{virtually become labeled samples}, which {substantially augment the training data}. 

%Compared with the conventional training methods (typically based on contrastive learning), knowledge distillation is preferable for two reasons. {Firstly, user's preferred items, whether interacted or not, will probably get highly-rated by the ranker (thanks to the high-precision of the ranker). As a result, such items will virtually become positive samples, which substantially augments the training data. Secondly, unlike the conventional methods where all the non-interacted items are equally treated as negative samples, user's real non-interested items will probably be lower rated by the ranker than the rest of non-interacted items, thus being penalized more when the retriever is trained. By doing so, the false negative issues can be effectively alleviated.}
}

{$\bullet$ \textbf{Training of ranker}. The ranker is trained by sampled softmax~\cite{bengio2008adaptive} on top of the \textbf{hard negative items} sampled by the retriever. Particularly, instead of working with the ``easy negatives'' which are randomly or heuristically sampled from the whole item set~\cite{rendle2009bpr,zhou2018deep,guo2017deepfm}, the ranker is iteratively trained to discriminate the true positive from the increasingly harder negatives as the retriever improves. Therefore, it prevents the ranker from converging too early to a limited discriminative capability. Besides, unlike the widely-used exact top-k sampling, we collect informative negative samples from the entire itemset based on the retriever; by doing so, it \textbf{alleviates the false negative issue} and \textbf{closes the gap between the training and inference stages}.}

{It's worth noting that the realization of the above training framework is non-trivial. Particularly, both retriever and ranker need to learn from the sampled items; however, the 
sampling operation on the retriever can be inefficient and biased. To overcome such challenges, a couple of technical designs are introduced.}
% The mutual reinforcement of the retriever and ranker will finally result in a highly competitive recommendation accuracy. To facilitate the optimized conduct of the proposed framework, a couple of technical designs are introduced for more effective sampling and knowledge distillation, respectively. 
Firstly, knowing that the directly sampling from retriever can be extremely time-consuming when dealing with a large itemset, we develop a \textbf{scalable and adaptive sampling strategy}, where the items favored by the retriever can be efficiently sampled in sublinear time with item size. Secondly, the direct knowledge distillation over the sampled items is biased and prone to inferior performances. To mitigate this problem, we propose a novel \textbf{asymptotic-unbiased estimation of KL divergence} for compensating the bias induced by item sampling. On top of this operation, the ranking order of items can be better aligned between the ranker and retriever.

We conduct comprehensive experimental studies over four benchmark datasets. According to the experiment results, the overall recommendation quality can be substantially improved by CoRR in comparison with the existing training methods. More detailed analysis further verifies CoRR's standalone effectiveness to both the retriever and the ranker, and its effectiveness as a model-agnostic training framework. The {contributions} of our work are summarized with the following points. 

\begin{itemize}[leftmargin=*]
	\item We present a novel training framework for deep recommender systems, where the retriever and ranker can be mutually reinforced for the effective cooperation. 
	\item Two critical techniques are introduced for the optimized conduct of CoRR: 1) the scalable and adaptive sampling strategy, {which enables the} efficient sampling from the retriever; 2) the asymptotic-unbiased estimation of KL divergence, as the objective of knowledge distillation, which better aligns the ranking order of items between the ranker and retriever, and {contributes to the retrieval of high-quality items}.
	%enabling items favored by retriever to be efficiently sampled
	\item We perform comprehensive experimental studies on four benchmark datasets, whose results verify CoRR's advantage against the existing training algorithms, and its standalone effectiveness to the retriever and ranker.
% 	deep recommenders, and its standalone effectiveness to the retriever and other jointly training strategies.
\end{itemize}

\section{Related Work}
This paper studied the cooperation between the ranker and retriever in deep recommender systems, to improve the recommendation quality of multi-stage systems. We first review closely related cascade ranking techniques, and then present negative sampling and knowledge distillation.
\subsection{Cascade Ranking}
Prior work on multi-stage cascade ranking usually assigned different rankers to each stage to achieve the desired trade-off collectively between efficiency and effectiveness~\cite{wang2011cascade}. They are different from each other in modeling the cost of each ranker~\cite{chen2017efficient,xu2013cost,xu2014classifier}. Recent work turned to directly optimizing cascade ranking models as a whole by gradient descent~\cite{gallagher2019joint} or identifying some bad cases with cascade ranking models to augment the training data~\cite{fan2019mobius}. Observing these cascading ranking systems do not consider the cooperation between rankers, the work~\cite{qin2022rankflow} suggested to optimize them jointly by letting cascade rankers provide supervised signals for each other in the cascading systems. However, the work is still limited from several critical perspectives, as aforementioned. This work aims to explore a better collaboration mechanism between cascade rankers.

%\subsection{Deep Learning for Recommendation}
%The earliest successful DL-based methods lie in modeling side information, such as textual data~\cite{huang2013learning,wang2015collaborative}, visual data~\cite{he2016vbpr}, knowledge graph~\cite{zhang2016collaborative} via deep learning techniques. However, behavior data is shallowly modeled based on inner product or Euclidean distance between user vector and item vector. Such deep models, which can be generalized to a two-tower similarity network, have an advantage in conveniently using the SOTA ANN algorithms. These methods have been extended by modeling feature interaction in an explicit or implicit way, whose representative works include Deep\&Wide~\cite{cheng2016wide}, NFM~\cite{he2017neuralfm}, DeepFM~\cite{guo2017deepfm}, XDeepFM~\cite{lian2018xdeepfm} and Deep\&Cross~\cite{wang2017deep}. Since behavior data is sequential in nature, therefore, it is intuitive to apply sequential deep learning methods for the sequential recommendation, where the representative works include SASRec~\cite{kang2018self} based on transformer, GRU4Rec based on GRU~\cite{hidasi2015session}, CASER based on CNN~\cite{tang2018personalized}. To refine user preferences, DIN~\cite{zhou2018deep} and DIEN~\cite{zhou2019deep} further incorporate complex feature interaction between recommended items and interaction history via the attention mechanism. Though recommendation performance can be improved in this way, recommendation efficiency is remarkably affected since the off-the-shelf ANN algorithms can not be applied anymore.

\subsection{Negative Sampling in RecSys}
Many methods sample negative items from static distributions, such as uniform distribution~\cite{rendle2009bpr,he2017neural} and popularity-based distribution~\cite{rendle2014improving}. To adapt to recommendation models, many advanced sampling methods have been proposed~\cite{rendle2014improving,hsieh2017colla,weston2010large,zhang2013optimizing,sun2019rotate,blanc2018adaptive,chen2021fast,lian2020personalized}. For example, AOBPR~\cite{rendle2014improving} transforms adaptive sampling into searching the item at a randomly-drawn rank. CML~\cite{hsieh2017colla} and WARP~\cite{weston2010large} draw negative samples for each positive by first drawing candidate items from uniform distribution and then selecting more highly-scored items than positive minus one as negative. Dynamic negative sampling (DNS)~\cite{zhang2013optimizing} picks a set of negative samples from the uniform distribution and then chooses the most highly-scored item as negative. Self-adversarial negative sampling~\cite{sun2019rotate} draws negative samples from the uniform distribution  but weights them with softmax-normalized scores. Kernel-based sampling~\cite{blanc2018adaptive} picks samples proportionally to a quadratic kernel in a divide and conquer way. Locality Sensitive Hashing (LSH) over randomly perturbed databases enables sublinear time sampling and LSH itself can generate correlated and unnormalized samples~\cite{spring2017new}. Quantization-based sampling~\cite{chen2021fast} decomposes the softmax-normalized probability via product quantization into the product of probabilities over clusters, such that sampling an item is in sublinear time. %MCMC has been used for approximately sampling items~\cite{yang2020understanding} by considering mixture of uniform over all items and uniform over $k$ nearest neighbors as proposal.
%IRGAN~\cite{wang2017irgan} distinguishes positive items from the randomly picked negatives according to the generator, but sampling is time-consuming. Though the generator (sampler) is simultaneously updated along with the discriminator, it is only trained by policy gradient (PG) to promote highly-probable items according to the discriminator. It does not only lack the supervision from the data but also suffers from PG-induced low training efficiency.

\subsection{Knowledge Distillation in RecSys}
KD~\cite{hinton2015distilling} provides a powerful model-agnostic framework for compressing a teacher model into a student model, by learning to imitate predictions from the teacher. Therefore, KD has been applied in RecSys for compressing deep recommenders. A pioneering work is Ranking Distillation (RD)~\cite{tang2018ranking}, where the top-$k$ recommendation from the teacher model is considered as position-weighted pseudo positive. The subsequent distillation methods improve the use of the top-$k$ recommendations, such as drawing a sample from the top-k items via ranking-based distribution~\cite{lee2019collaborative}, its mixture with uniform distribution~\cite{kang2020rrd} and rank discrepancy-aware distribution~\cite{kweon2021bidirectional}. In addition to teacher's prediction, the latent knowledge in the teacher can be distilled into the student via hint regression~\cite{kang2020rrd} and topology distillation~\cite{kang2021topology}. %KD has been applied in several jointly methods in recommendation, such as the topk-loss in RankFlow~\cite{qin2022rankflow}.  In contrast to them, our method does not depend on the top-k results of the retriever at all. 

\section{Cooperative Retriever and Ranker}
In this section, we first overview the training of the CoRR on a training dataset $\mathcal{D}=\{(c,k)\}$, which consists of pairs of a context $c$ and a positive item $k$. The context indicates all information except items, such as user, interaction history, time and location. Following that, we elaborate on how to train the ranker with the retriever based on hard negative sampling and how to train the retriever with the ranker based on knowledge distillation. 
\subsection{Overview}
For the sake of scalable recommendation, deep recommender relies on the collaboration of two models: the retrieval model (retriever) and the ranking model (ranker). The retriever targets on selecting a small set of potentially positive items from the whole items with high efficiency. Typically, the retriever is represented by $M_\theta(i,c)=\text{sim}\big(E_{\theta_1}(i),E_{\theta_2}(c)\big)$, i.e., the similarity (e.g. Euclidean distance, inner product and cosine similarity) between item embedding $E_{\theta_1}(i)$ and context embedding $E_{\theta_2}(c)$. Therefore, we can use off-the-shelf ANNs, such as FAISS~\cite{JDH17} and SCANN~\cite{guo2020accelerating}, for retrieving the most similar items as candidates in sublinear time. The ranker aims to identify the best items from the retrieval results with high precision, and is usually represented by expressive yet time-consuming networks $R_\phi(i,c)$ directly taking an item $i$ and a context $c$ as inputs. 

Due to time-consuming computation in the ranker, in most cases, traditional training methods learn to discriminate positive from randomly picked negatives. In spite of simplicity, the ranker is trained independently to the retriever. In some cases, the ranker is trained by discriminating positives from the top-k retrieval results from the retriever. In spite of establishing a connection with the retriever, the ranker easily suffers from the false negative issue, since potential positives are definitely included in the top-k retrieval results. Moreover, these negatives may introduce a large bias to gradient, such that the optimization may converge slowly and even be easily stuck into a local optimum. To this end, we suggest to optimize the ranker w.r.t the sampled log-softmax~\cite{bengio2008adaptive}, an asymptotic-unbiased estimation of log-softmax based on importance sampling. The sampled log-softmax relies on randomly-drawn samples from a proposal distribution. To connect the ranker with the retriever, we propose to \textbf{sample negatives from the retriever based on an adaptive and scalable strategy}.

\begin{figure}[t]
	\centering
	\includegraphics[width=0.8\linewidth]{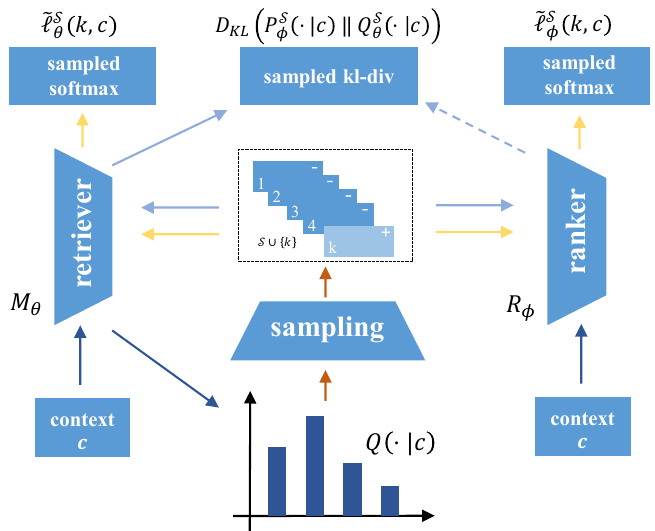}
	\vspace{-0.3cm}
	\caption{The framework of CoRR. The context $c$ indicates all information except items. 
% 	$Y(\cdot)$ indicates the static proposal and 
	$Q(\cdot|c)$ indicates the proposal distribution given context $c$. The dotted line means the gradient is stopped.} \label{fig:corrframework}
	\Description[The framework of CoRR.]{Fully described in the text in Section 3.1.}
	\vspace{-.5cm}
\end{figure}

 In addition to supervision from data, the training of the retriever can also be guided by the ranker based on knowledge distillation since the ranker is assumed more precise at ranking items. Regarding supervision from recommendation data, the sampled log-softmax objective is also exploited for optimizing the retriever with the same proposal in the ranker optimization. Regarding knowledge distillation from the ranker, most prior methods distill knowledge in top-k results from the ranker~\cite{tang2018ranking,lee2019collaborative,kang2020rrd}, but the top-k ranking results are time-consuming for the ranker. In this paper, we distill the ranking order information from the ranker's predictions, by directly aligning softmax-normalized predictions between the ranker and retriever based on the KL divergence. To reduce the time cost of optimization, we propose \textbf{an asymptotic-unbiased estimation for the KL divergence}, named \textbf{sampled KL divergence}.

The retriever and ranker are simultaneously trained with a unified workflow, where both models can be mutually reinforced. As the training progresses, the ranker becomes increasingly precise, which in return provides more informative supervision signals for the retriever; meanwhile, as the retriever improves, negative samples will become increasingly harder, which contributes to a higher discriminative capability of the ranker. The overall framework is illustrated in Figure~\ref{fig:corrframework}. The overall procedure can be referred to in Algorithm~\ref{alg:corr}.

\subsection{Training Ranker with Retriever}
\subsubsection{\textbf{Loss Function}}
In this paper, we optimize the ranker w.r.t sampled log-softmax, which is an asymptotic-unbiased estimation of log-softmax. The use of log-softmax is motivated by its close relationship with the logarithm of both Mean Reciprocal Rank (MRR) and Normalized Discounted Cumulative Gain (NDCG)~\cite{bruch2019analysis} as well as the excellent recommendation performance~\cite{liang2018variational}. Assuming the positive item is $k$ at a context $c$, the log-softmax objective w.r.t $k$ is formulated as follows:
\begin{displaymath}
\begin{aligned}
    \ell(k,c) & = R_\phi(k,c) - \log \Big(\sum_i \exp\big(R_\phi(i,c)\big)\Big)\\
    & = - \log \Big(\sum_i \exp\big(R_\phi(i,c) - R_\phi(k,c)\big)\Big) \\
    &\le -\log \Big(\sum_i \mathbb{I}_{\ge 0}\big( R_\phi(i,c) - R_\phi(k,c) \big) \Big)\\
    &=\log \Big(\frac{1}{\text{rank}_R(k,c)}\Big) = \log \text{MRR}\\
\end{aligned}
\end{displaymath}
where the inequality holds due to $\exp(\cdot) \ge \mathbb{I}_{\ge 0}(\cdot)$. 

\setlength{\textfloatsep}{0.1cm}
\begin{algorithm}[t]
	\KwIn{Training Data $\mathcal{D}$}
	\KwOut{Retriever $M_\theta$ and Ranker $R_\phi$ }
	\Repeat{Convergence or reaching maxepoch}
	{
		\ForEach{$(c,k) \in \mathcal{D}$}
		{
			$\mathcal{S} \gets $ Draw $n$ samples according to ${Q}(\cdot|c)$\;
% 			$M_\theta(\cdot,c) \gets $ Forward-Inference($M_\theta$, $\mathcal{S}\cup \{k\}$) \;
% 			$Q_{\mathcal{C}\cup \{k\}}(\cdot|c) \gets \text{softmax}\Big(M_\theta(\cdot,c)/T - \log Y(\cdot)\Big)$\;
% 			$\mathcal{S} \gets$ Draw $L$ samples according to $Q_{\mathcal{C}\cup \{k\}}(\cdot|c)$\;
			Update $M_\theta$ w.r.t. $\tilde{\ell}_\theta^\mathcal{S}(k,c)+D_{KL}(P_\phi^\mathcal{S}(\cdot|c)\parallel {P}_\theta^\mathcal{S}(\cdot|c))$\;
			Update $R_\phi$ w.r.t.
			$\tilde{\ell}_\phi^\mathcal{S}(k,c)$\;
		}
	}
	\caption{Cooperative Retriever and Ranker}\label{alg:corr}
\end{algorithm}
\setlength{\floatsep}{0.1cm}

However, it is computationally challenging to optimize the log-softmax, since the gradient computation scales linearly with the number of items, i.e.,
\begin{displaymath}
    \nabla \ell_\phi(k,c) = \nabla R_\phi(k,c) - \mathbb{E}_{i\sim P_\phi(\cdot|c)} [ \nabla R_\phi(i,c)]
\end{displaymath}
where $P_\phi(\cdot|c)$ represents categorical probability distribution over the whole items $\mathcal{I}$ with the parameter $\text{softmax}(R_\phi(\cdot,c))$, i.e., $P_\phi(j|c)=\frac{\exp(R_\phi(j,c)}{\sum_{i\in \mathcal{I}}\exp(R_\phi(i,c))}$. 
To improve the efficiency, importance sampling was used for approximation by drawing $L$ samples $\mathcal{S}=\{o_1, \cdots, o_L\}$ from the proposal distribution ${Q}(\cdot|c)$.  That is,
\begin{displaymath}
    \nabla \ell_\phi(k,c) \approx \nabla R_\phi(k,c) - \sum_{i\in \mathcal{S}\cup \{k\}} w(i,c) \nabla R_\phi(i,c)
\end{displaymath}
where $w(i,c)=\frac{\exp\big(R_\phi(i,c)-\log \tilde{Q}(i|c)\big)}{\sum_{j\in \mathcal{S}\cup \{k\}} \exp\big(R_\phi(j,c) - \log \tilde{Q}(j|c)\big)}$. It is easy to verify that this gradient can be derived from the following sampled log-softmax.
\begin{equation}
	 \tilde{\ell}_\phi^\mathcal{S}(k,c) =\log \frac{\exp\big(R_\phi(k,c) - \log \tilde{Q}(k|c)\big)}{\sum_{i\in \mathcal{S}\cup \{k\}} \exp\big(R_\phi(i,c) - \log \tilde{Q}(i|c)\big)}
	\label{eq:sample_logsoftmax}
\end{equation}
where $\tilde{Q}(\cdot|c)$ is the unnormalized ${Q}(\cdot|c)$. Note that both positive item $k$ and sampled itemset $\mathcal{S}$ are used in these formulations. This is because it is necessary to guarantee negativeness of the sampled log-softmax like log-softmax. To connect the ranker with the retriever, we propose to construct the proposal distribution from the retriever $M_\theta$ in this paper. In the next part, we will elaborate on how to efficiently sample items based on the retriever.

\subsubsection{\textbf{Sampling from the Retriever}} \label{sec:resample}
Prior work considered the top-k retrieval results as negative, they suffer from the false negative issues since the top-k results include both hard negatives and potential positives. In this paper, we propose to construct the proposal distribution from the retriever $M_\theta$, i.e., $Q(i|c) = P_\theta(i|c)= \frac{\exp\big(M_\theta(i,c) /T\big)}{\sum_{j\in \mathcal{I}} \exp\big(M_\theta(j,c)/T\big)}$, where the coefficient $T$ controls the balance between hardness and randomness of negative samples. When $T\rightarrow \infty$, sampling approaches completely random; when $T\rightarrow 0$, it becomes fully deterministic, being equivalent to argmax. In this case, the sampled log-softmax can be reformulated as $$\tilde{\ell}_\phi^\mathcal{S}(k,c)=\log \frac{\exp\big(R_\phi(k,c) - M_\theta(k,c)/T\big)}{\sum_{i\in \mathcal{S}\cup \{k\}} \exp\big(R_\phi(i,c) - M_\theta(i,c)/T\big)}.$$

% In this part, we propose a scalable and adaptive sampler for approximately sampling from $Q(i|c)$, which consists of two steps:
% \begin{itemize}[leftmargin=*]
% 	\item The first step is to draw a comparatively large pool (of size $n$) of candidate samples from a static categorical distribution $Y(\cdot)$, such as uniform and popularity-based distribution.
% 	\item The second step is to draw a small size of ($L<n$) samples from the candidate pool with replacement based on the inference results in the retriever $M_\theta$. 
% \end{itemize}
However, it is time-consuming to draw samples from $P_\theta(i|c)$ while sampling efficiency is important for training model. In the part, we introduce a scalable and adaptive sampler, which enables to draw samples from $P_\theta(i|c)$ in sublinear time. Basically, the sampler first builds a multi-level inverted index on the item embeddings in retrieval model with the inner product similarity and then draws negative samples based on the index. Regarding index building, each item embedding is first evenly split into two subvectors, and in each subspace, items are clustered with the K-means. Each item can then be approximated by concatenation of the corresponding cluster centers. Particularly, item $i$'s embedding $E_{\theta_1}(i)\approx \mathbf{w}_{k(i)}^1\oplus \mathbf{w}_{k(i)}^2$, where $k(i)$ indicates the cluster assignment of the item $i$ while $\mathbf{w}_{k(i)}^1$ and $\mathbf{w}_{k(i)}^2$ denote cluster centers of the item $i$. Let $\Omega_k^1$ and $\Omega_k^2$ be the item set belong to the cluster $k$ in the first and second subspace respectively and split query vector $\mathbf{z}_c=E_{\theta_2}(c)$ of the context $c$ into two parts, i.e., $\mathbf{z}_c=\mathbf{z}^1_c\oplus \mathbf{z}^2_c$. Negative sampling can then be decomposed into the following three steps:
\begin{itemize}[leftmargin=*]
    \item \textbf{Sampling a cluster $k^1$ in the first subspace.} The sampling probability of the cluster $k$ is defined as $P(k) = \frac{\psi_k \exp(\langle \mathbf{z}^1_c, \mathbf{w}_k^1\rangle)}{\sum_{k'} \psi_{k'} \exp(\langle \mathbf{z}^1_c, \mathbf{w}_{k'}^1\rangle)}$, where $\psi_k = \sum_{k'} \omega_{k,k'} \exp(\langle \mathbf{z}^2_c, \mathbf{w}_{k'}^2\rangle)$ while $\omega_{k,k'}=\lvert\Omega_k^1 \cap \Omega_{k'}^2\rvert$.
    \item \textbf{Sampling a cluster $k^2$ in the second subspace conditional on $k^1$.} The sampling probability of the cluster $k$ is defined as $P(k|k^1) = \frac{\omega_{k^1,k} \exp(\langle \mathbf{z}^2_c, \mathbf{w}_{k}^2\rangle)}{\sum_{k'} \omega_{k^1,k'} \exp(\langle \mathbf{z}^2_c, \mathbf{w}_{k'}^2\rangle)}$.
    \item \textbf{Sampling an item uniformly within the intersection item set $\Omega_{k_1}^1 \cap \Omega_{k_2}^2$.}
\end{itemize}

\textbf{Remarks on Sampling Effectiveness}: In spite of being decomposed, this sampling procedure actually corresponds to sampling from $\hat{Q}(i|c) = \frac{\exp(\langle \mathbf{z}_c, \mathbf{w}_{k(i)}^1\oplus \mathbf{w}_{k(i)}^2 \rangle)}{\sum_j \exp(\langle \mathbf{z}_c, \mathbf{w}_{k(j)}^1\oplus \mathbf{w}_{k(j)}^2 \rangle)}$. Thanks to the bounded divergence from above between $\hat{Q}(i|c)$ and $Q(i|c)$, the sampling effectiveness can be guaranteed~\cite{chen2021fast}, depending on the residual error of clustering. When the residual error is smaller, approximate sampling is more closely to exact sampling.

\textbf{Remarks on Sampling Efficiency}:
Since $\omega_{k,k'}$ is query independent, it can be precomputed after clustering. The overall time complexity of sampling $T$ items is $\mathcal{O}(Kd+K^2+T)$, where $d$ is the embedding dimension and $K$ is the number of clusters in K-means.

\subsection{Training Retriever with Ranker}
Although the retriever has been used for training the ranker, the gradient from the ranker's objective should be stopped, since the retriever is only used for providing negative information for the ranker. The training of the retriever $M_\theta(i,c)$ is then guided by supervision loss from the recommendation data and a distillation loss from the ranker.
\subsubsection{\textbf{Supervision Loss}}
Since the retriever concerns about the recall performance of retrieval results, it should optimize the ranking performance. Similar to the ranker, the retriever also exploits the sampled log-softmax for optimization as long as simply replacing $R_\phi(i|c)$ with $M_\theta(i,c)$. We also use the sampler in Section~\ref{sec:resample} as the proposal, since it can greatly reduce the cost of forward inference and back-propagation in the retriever by using a few sampled items. Therefore, the objective function is then represented as 
\begin{equation}
	\tilde{\ell}_\theta^\mathcal{S}(k,c) =\log \frac{\exp\big(M_\theta(k,c) - \log \tilde{Q}(k|c)\big)}{\sum_{i\in \mathcal{S}\cup \{k\}} \exp\big(M_\theta(i,c) - \log \tilde{Q}(i|c)\big)}
\end{equation} 
\subsubsection{\textbf{Distillation Loss}} \label{sec:distillation}
The high efficiency of the retriever comes at the cost of limited expressiveness. Knowing that the ranker is more precise, the knowledge distillation may provide substantial weak-supervision signals to alleviate the sparsity of training data. Therefore, in this section, we design the knowledge distillation loss from the ranker. Without concentrating on specific recommenders, we consider not distilling latent knowledge but the predictions from the ranker. To avert the use of the top-k results from the ranker in prior work, we follow the pioneering work of KD~\cite{hinton2015distilling} to directly match softmax-normalized predictions between the ranker and the retriever via KL divergence, that is,
\begin{equation}
	D_{KL}\Big(P_\phi(\cdot|c)\parallel P_\theta(\cdot|c)\Big) = \sum_{i\in \mathcal{I}} P_\phi(i|c) \log \frac{P_\phi(i|c)}{P_\theta(i|c)},
	\label{eq:kl_loss}
\end{equation}
where $P_\phi(\cdot|c)$ and $P_\theta(\cdot|c)$ are probabilities induced by the ranker and retriever, respectively. However, it is impracticable to directly compute KL divergence and its gradient, since it scales linearly with the number of items w.r.t each context. To this end, we propose an asymptotic-unbiased estimation for the KL divergence for speeding up forward inference and backward propagation. In particular, first denote by $\mathcal{S}=\{o_1,o_2,\cdots, o_L\}$ the sample set drawn from the proposal ${Q}(\cdot|c)$ and define ${P}_\phi^\mathcal{S}(j|c)=\frac{\exp(\tilde{R}_\phi(j,c))}{\sum_{i\in \mathcal{S}}\exp(\tilde{R}_\phi(i,c))}$ and ${P}_\theta^\mathcal{S}(j|c)=\frac{\exp(\tilde{M}_\theta(j,c))}{\sum_{i\in \mathcal{S}}\exp(\tilde{M}_\theta(i,c))}$, where $\tilde{R}_\phi(i,c) =R_\phi(i,c) - \log \tilde{Q}(i|c)$ and $\tilde{M}_\theta(i,c) = M_\theta(i,c) - \log \tilde{Q}(i|c)$. $D_{KL}\Big({P}_\phi^\mathcal{S}(\cdot |c)\parallel {P}_\theta^\mathcal{S}(\cdot |c)\Big)$ is the asymptotic-unbiased estimation for $D_{KL}\Big(P_\phi(\cdot|c)\parallel P_\theta(\cdot|c)\Big)$ according to the following theorem.
\begin{theorem}
	$D_{KL}\Big({P}_\phi^\mathcal{S}(\cdot |c)\parallel {P}_\theta^\mathcal{S}(\cdot |c)\Big)$ converges to $D_{KL}\Big(P_\phi(\cdot|c)\parallel P_\theta(\cdot|c)\Big)$ with probability 1 when $L=\lvert \mathcal{S} \rvert \rightarrow \infty$.
	\label{thm:kl_convergence}
\end{theorem}

When used for learning the retriever, the sampled KL divergence is based on the sampler in Section~\ref{sec:resample}. Below we discuss some special cases of samplers for further understanding its generality.

\subsubsection{\textbf{Special Cases}} In this part, we investigate two special proposals: ${Q}(\cdot|c)=P_\theta(\cdot|c)=\text{softmax}(M_\theta(\cdot,c))$ and ${Q}(\cdot|c)$ is uniform.
 
$\bullet$ \textbf{${Q}(\cdot|c)=P_\theta(\cdot|c)$}: In this case the ranker is optimized by the sampled log-softmax $\tilde{\ell}(k,c)=\log \frac{\exp(\Delta_k^c)}{\sum_{i\in \mathcal{S}} \exp(\Delta_{i}^c)}$, where $\Delta_k^c=R_\phi(k,c) - M_\theta(k,c)$. The distillation loss could be simplified as follows:
\begin{corollary}
	If each sample in $\mathcal{S}$ is drawn according to $P_\theta(\cdot|c)$, concatenating $\{\Delta_{i}^c|i\in \mathcal{S}\}$ as a vector $\boldsymbol{\Delta}^c$ of length $L$, then $D_{KL}(P_\phi\parallel P_\theta)$ can be asymptotic-unbiasedly estimated by $\log L -  H({\normalfont \text{softmax}}(\boldsymbol{\Delta}^c)) $, where $H(p)$ is the Shannon entropy of categorical distribution parameterized by $p$. 
	\label{thm:kl_corollary}
\end{corollary}
The proof is provided in the appendix. The corollary indicates that when exactly drawing samples from $P_\theta(\cdot|c)$, minimizing the KL divergence is equivalent to maximizing the entropy of a categorical distribution. The distribution is parameterized by softmax-normalized prediction differences between the ranker and retriever.

$\bullet$ \textbf{${Q}(\cdot\vert c)$ is uniform}: In this case, the ranker is optimized by the sampled log-softmax $\tilde{\ell}(k,c)=\log \frac{\exp(R_\phi(k,c))}{\sum_{i\in \mathcal{S}} \exp(R_\phi(i,c))}$. And ${P}_\mathcal{S}(\cdot|c)$ and ${Q}_\mathcal{S}(\cdot|c)$ could be simplified: ${P}_\phi^\mathcal{S}(j|c)=\frac{\exp(R_\phi(j,c))}{\sum_{i\in \mathcal{S}}\exp(R_\phi(i,c))}$ and ${P}_\theta^\mathcal{S}(j|c)=\frac{\exp(M_\theta(j,c))}{\sum_{i\in \mathcal{S}}\exp(M_\theta(i,c))}$. Minimizing the KL divergence between them is equivalent to matching the ranking order on the randomly sample set $\mathcal{S}$ between the ranker and retriever.

\section{Experiments}
Experiments are conducted to verify the effectiveness of the proposed CoRR, by answering the following questions:
\begin{itemize}
    \item[\textbf{RQ1:}] Does CoRR outperforms conventional DRS and the existing joint training methods?
    \item[\textbf{RQ2:}] Could the adaptive sampler generate higher-quality negative items than the exact top-k sampling to help training?
    \item[\textbf{RQ3:}] Does the ranking-order preserving distillation loss improve the retriever?
    % \item[\textbf{RQ4:}] Is the pipeline of CoRR effective for various retriever and ranker pairs and for different tasks?
\end{itemize}
Since CoRR is model-agnostic, to demonstrate the effectiveness of CoRR, we apply the framework for both general recommendation and sequential recommendation in this paper.

% By observing the representativeness and importance of sequential recommendation in recommender systems, we only evaluate it in sequential recommendation tasks in this paper.

\subsection{Experimental Settings}
\subsubsection{Datasets}
As shown in Table~\ref{tab0}, we evaluate our method on four real-world datasets. The datasets are from different domains and platforms, and they vary significantly in size and sparsity.
Gowalla dataset contains users' check-in data at locations at different times. Taobao dataset is a big industrial dataset collected by Alibaba Group, which contains user behaviors including click, purchase, adding item to shopping cart, and item favoring. We select the largest subset which contains click behaviors. The Amazon dataset is a subset of product reviews for Amazon Electronics. MovieLens dataset is a classic movie rating dataset, in which ratings range from 0.5 to 5. We choose a subset with 10M interactions to conduct experiments. Then we filter out users and items (locations/products/movies) less than 10 interactions for all datasets.

For general recommendation task, the behavior history of each user is splited in to train/valid/test by ratio 0.8/0.1/0.1. For sequential recommendation task, given the behavior history of a user is $(i_1,i_2,\cdots,i_k,\cdots,i_n)$, the goal is to predict the $(k+1)$-th items using the first $k$ items. In all experiments, we generate the training set with $k=1,2,\cdots,n-3$ for all users, and we predict the next one given the first $n-2$ and $n-1$ items in the valid and test set respectively. Besides, we set the max sequence length to 20 for the user behavior sequence in all datasets.

\begin{table}[t]
\caption{Dataset Statistics.}
%\vspace{-1em}
\renewcommand\arraystretch{0.9}
\begin{tabular}{c|rrr}
\toprule % \hline
          & \#users & \#items & \#interactions \\ \midrule % \hline
Amazon    & 9,280  & 6,066   & 158,979        \\
Gowalla   & 29,859   & 40,989  & 1,027,464        \\
MovieLens & 66,958  & 10,682   & 5,857,041       \\  %\hline
Taobao    & 941,853  & 1101,236  & 63,721,355       \\
\bottomrule
\end{tabular}
\label{tab0}
%\vspace{-0.3cm}
\end{table}

% main table: compare with baselines
\begin{table*}[t]
% \small
\centering
\caption{Comparisons with baselines($\times 10^{-2}$). $\triangle$,$\blacktriangle$ indicate the improvements of CoRR over the best results of baselines are statistically significant for $p<0.05$,  $p< 0.001$ based on t test.}
%\vspace{-0.3cm}

% \resizebox{\textwidth}{5cm}{
\renewcommand\arraystretch{0.8}
\begin{tabular}{l|l|cccc|cccc}
    \toprule
    Dataset & Metrics & BPR & NCF & LogisticMF & DSSM & Independent & ICC & {RankFlow} & {CoRR} \\ \midrule
        \multirow{3}{*}{Amazon} & \multicolumn{1}{l|}{Recall@10}
        & $ 4.86 $ & $ 4.51 $ & $ 3.74 $ & $ 4.55 $ & $ 4.97 $ & $ 4.88 $ & $ 5.06 $ & $ \mathbf{5.76}^{\blacktriangle} $ \\
        \multicolumn{1}{l|}{}       & \multicolumn{1}{l|}{NDCG@10}
        & $ 2.72 $ & $ 2.52 $ & $ 2.03 $ & $ 2.52 $ & $ 2.68 $ & $ 2.74 $ & $ 2.82 $ & $ \mathbf{3.22}^{\blacktriangle} $ \\
        \multicolumn{1}{l|}{}       & \multicolumn{1}{l|}{MRR@10}
        & $ 2.51 $ & $ 2.29 $ & $ 1.80 $ & $ 2.28 $ & $ 2.43 $ & $ 2.37 $ & $ 2.63 $ & $ \mathbf{3.04}^{\blacktriangle} $ \\
        \midrule \multirow{3}{*}{Gowalla} & \multicolumn{1}{l|}{Recall@10}
        & $ 7.78 $ & $ 8.33 $ & $ 6.70 $ & $ 6.30 $ & $ 8.24 $ & $ 9.08 $ & $ 9.35 $ & $ \mathbf{10.52}^{\blacktriangle} $ \\
        \multicolumn{1}{l|}{}       & \multicolumn{1}{l|}{NDCG@10}
        & $ 5.74 $ & $ 5.69 $ & $ 4.21 $ & $ 4.32 $ & $ 6.17 $ & $ 6.18 $ & $ 6.34 $ & $ \mathbf{7.49}^{\blacktriangle} $ \\
        \multicolumn{1}{l|}{}       & \multicolumn{1}{l|}{MRR@10}
        & $ 7.91 $ & $ 7.41 $ & $ 4.99 $ & $ 5.88 $ & $ 7.96 $ & $ 8.14 $ & $ 8.65 $ & $ \mathbf{10.17}^{\blacktriangle} $ \\
        \midrule \multirow{3}{*}{MovieLens} & \multicolumn{1}{l|}{Recall@10}
        & $ 18.23 $ & $ 18.63 $ & $ 10.79 $ & $ 14.83 $ & $ 19.54 $ & $ 20.22 $ & $ 20.72 $ & $ \mathbf{22.44}^{\triangle} $ \\
        \multicolumn{1}{l|}{}       & \multicolumn{1}{l|}{NDCG@10}
        & $ 16.54 $ & $ 16.82 $ & $ 9.44 $ & $ 13.88 $ & $ 17.49 $ & $ 17.96 $ & $ 18.73 $ & $ \mathbf{21.59}^{\blacktriangle} $ \\
        \multicolumn{1}{l|}{}       & \multicolumn{1}{l|}{MRR@10}
        & $ 25.98 $ & $ 26.22 $ & $ 15.51 $ & $ 22.59 $ & $ 27.74 $ & $ 27.76 $ & $ 28.73 $ & $ \mathbf{33.70}^{\blacktriangle} $ \\
        \midrule \multirow{3}{*}{TaoBao} & \multicolumn{1}{l|}{Recall@10}
        & $ 0.87 $ & $ 1.87 $ & $ 0.51 $ & $ 1.66 $ & $ 2.16 $ & $ 2.19 $ & $ 2.63 $ & $ \mathbf{2.79}^{\triangle} $ \\
        \multicolumn{1}{l|}{}       & \multicolumn{1}{l|}{NDCG@10}
        & $ 0.67 $ & $ 0.94 $ & $ 0.25 $ & $ 0.83 $ & $ 1.04 $ & $ 1.12 $ & $ 1.24 $ & $ \mathbf{1.43}^{\triangle} $ \\
        \multicolumn{1}{l|}{}       & \multicolumn{1}{l|}{MRR@10}
        & $ 0.62 $ & $ 0.65 $ & $ 0.17 $ & $ 0.58 $ & $ 0.72 $ & $ 0.79 $ &	0.88 & $ \mathbf{1.02}^{\blacktriangle} $ \\

    \bottomrule \toprule
    Dataset & Metrics & {GRU4Rec} & {BERT4Rec} & {Caser} & {SASRec} & {Independent} & {ICC} & {RankFlow} & {CoRR} \\ \midrule
        \multirow{3}{*}{Amazon} & \multicolumn{1}{l|}{Recall@10}
        & $ 4.98 $ & $ 5.18 $ & $ 4.74 $ & $ 5.18 $ & $ 5.26 $ & $ 4.34 $ & $ 5.33 $ & $ \mathbf{5.84}^{\triangle} $ \\
        \multicolumn{1}{l|}{}       & \multicolumn{1}{l|}{NDCG@10}
        & $ 2.61 $ & $ 2.81 $ & $ 2.46 $ & $ 2.63 $ & $ 2.69 $ & $ 2.34 $ & $ 2.78 $ & $ \mathbf{3.07}^{\triangle} $ \\
        \multicolumn{1}{l|}{}       & \multicolumn{1}{l|}{MRR@10}
        & $ 1.89 $ & $ 2.00 $ & $ 1.77 $ & $ 1.85 $ & $ 1.94 $ & $ 1.73 $ & $ 2.06 $ & $ \mathbf{2.23}^{\triangle} $ \\
        \midrule \multirow{3}{*}{Gowalla} & \multicolumn{1}{l|}{Recall@10}
        & $ 8.43 $ & $ 13.61 $ & $ 7.88 $ & $ 9.29 $ & $ 12.50 $ & $ 9.28 $ & $ 12.68 $ & $ \mathbf{14.46}^{\triangle} $ \\
        \multicolumn{1}{l|}{}       & \multicolumn{1}{l|}{NDCG@10}
        & $ 4.75 $ & $ 7.74 $ & $ 4.67 $ & $ 5.42 $ & $ 7.06 $ & $ 5.28 $ & $ 7.09 $ & $ \mathbf{8.31}^{\triangle} $ \\
        \multicolumn{1}{l|}{}       & \multicolumn{1}{l|}{MRR@10}
        & $ 3.64 $ & $ 5.95 $ & $ 3.69 $ & $ 4.24 $ & $ 5.41 $ & $ 4.07 $ & $ 5.82 $ & $ \mathbf{6.44}^{\triangle} $ \\
        \midrule \multirow{3}{*}{MovieLens} & \multicolumn{1}{l|}{Recall@10}
        & $ 15.54 $ & $ 14.67 $ & $ 16.78 $ & $ 15.02 $ & $ 16.89 $ & $ 17.27 $ & $ 17.59 $ & $ \mathbf{18.46}^{\blacktriangle} $ \\
        \multicolumn{1}{l|}{}       & \multicolumn{1}{l|}{NDCG@10}
        & $ 7.78 $ & $ 7.52 $ & $ 8.55 $ & $ 7.53 $ & $ 8.97 $ & $ 9.23 $ & $ 9.33 $ & $ \mathbf{9.83}^{\triangle} $ \\
        \multicolumn{1}{l|}{}       & \multicolumn{1}{l|}{MRR@10}        & $ 5.45 $ & $ 5.37 $ & $ 6.07 $ & $ 5.09 $ & $ 5.96 $ & $ 6.21 $ & $ 6.60 $ & $ \mathbf{7.25}^{\blacktriangle} $ \\
        \midrule \multirow{3}{*}{Taobao}    & \multicolumn{1}{l|}{Recall@10}
        & $ 0.49 $ & $ 0.68 $ & $ 0.37 $ & $ 0.30 $ & $ 0.89 $ & $ 0.88 $ & $ 0.94 $ & $ \mathbf{1.16}^{\blacktriangle} $ \\
        \multicolumn{1}{l|}{}       & \multicolumn{1}{l|}{NDCG@10}
        & $ 0.42 $ & $ 0.61 $ & $ 0.30 $ & $ 0.26 $ & $ 0.76 $ & $ 0.73 $ &	0.89 & $ \mathbf{1.01}^{\blacktriangle} $ \\
        \multicolumn{1}{l|}{}       & \multicolumn{1}{l|}{MRR@10}
        & $ 0.88 $ & $ 1.14 $ & $ 0.65 $ & $ 0.57 $ & $ 1.58 $ & $ 1.55 $ &	1.96 & $ \mathbf{2.13}^{\blacktriangle} $ \\
        \bottomrule
    \end{tabular}
% }
\label{tab1}
\vspace{-0.1cm}
\end{table*}

% extend retriever and ranker
\begin{table*}[h]
\small
\caption{Extensive Retriever and Ranker. ($\times 10^{-2}$)}
%\vspace{-0.3cm}
\setlength\tabcolsep{2pt}
\renewcommand\arraystretch{0.8}
\begin{tabular}{l|l|ccc|ccc||ccc|ccc}
\toprule
\multirow{2}{*}{Dataset} &
  \multirow{2}{*}{Metric} &
  \multicolumn{3}{c}{MF+DeepFM} &
  \multicolumn{3}{c||}{DSSM+DCN} &
  \multicolumn{3}{c}{SASRec+DIN} &
  \multicolumn{3}{c}{Caser+BST} \\ \cmidrule{3-8} \cmidrule{9-14}
                           & & Independent & RankFlow & CoRR   & Independent & RankFlow & CoRR   & Independent & RankFlow & CoRR   & Independent & RankFlow & CoRR   \\ \midrule
\multirow{3}{*}{Amazon}    & Recall@10 & 4.97 & 5.06 & \textbf{5.76} & 4.28 & 4.44 & \textbf{4.72} & 5.26 & 5.33 & \textbf{5.84} & 4.77 & 4.79 & \textbf{5.03} \\
                           & NDCG@10   & 2.68 & 2.82 & \textbf{3.22} & 2.36 & 2.45 & \textbf{2.68} & 2.69 & 2.78 & \textbf{3.07} & 2.57 & 2.60 & \textbf{2.72} \\
                           & MRR@10    & 2.43 & 2.63 & \textbf{3.04} & 2.12 & 2.24 & \textbf{2.45} & 1.94 & 2.06 & \textbf{2.23} & 1.83 & 1.87& \textbf{1.98} \\ \midrule
\multirow{3}{*}{Gowalla}   & Recall@10 & 8.24 & 9.35 & \textbf{10.52} & 7.01 & 8.12 & \textbf{9.69} & 12.50 & 12.68 & \textbf{14.46} & 10.99 & 10.68 & \textbf{13.90} \\
                           & NDCG@10   & 6.17 & 6.56 & \textbf{7.54}  & 5.17 & 6.10 & \textbf{6.73} & 7.06 & 7.09 & \textbf{8.31}  & 6.03 & 6.11 & \textbf{8.27}  \\
                           & MRR@10    & 7.96 & 8.94 & \textbf{10.26} & 7.28 & 7.94 & \textbf{8.97} & 5.41 & 5.82 & \textbf{6.44}  & 4.53 & 4.73 & \textbf{6.55}  \\ \midrule
\multirow{3}{*}{MovieLens} & Recall@10 & 19.54 & 20.72 & \textbf{22.44} & 18.51 & 20.37 & \textbf{22.94} & 16.89 & 17.59 & \textbf{18.46} & 17.48 & 16.54 & \textbf{23.16} \\
                           & NDCG@10   & 17.49 & 18.73 & \textbf{22.26} & 18.01 & 20.10 & \textbf{22.08} & 8.97 & 9.33 & \textbf{9.83}  & 8.84 & 8.33 & \textbf{12.59} \\
                           & MRR@10    & 27.84 & 28.73 & \textbf{34.07} & 29.02 & 31.53 & \textbf{34.24} & 5.96 & 6.60 & \textbf{7.25}  & 6.24 & 6.22 & \textbf{9.39}  \\
\bottomrule
\end{tabular}
\label{tab6:extend}
\vspace{-0.2cm}
\end{table*}

\subsubsection{Metric}
Three common top-k metrics are used in evaluation, NDCG~\cite{weimer2007cofi}, Recall~\cite{hu2008collaborative, wang2011collaborative} and MRR. Recall@k represents the proportion of cases when the target item is among the top k items. NDCG@k gives higher weights on higher ranks. MRR@k represents the average of reciprocal ranks of target items. A greater value of these metrics indicates better performance. 

\subsubsection{Implementation Details}\label{sec:details}
In this paper, MF (Matrix Factorization)~\cite{5197422, rendle2009bpr} and DeepFM~\cite{guo2017deepfm} are selected as retriever and ranker respectively in general recommendation. In sequential recommendation, we set SASRec~\cite{kang2018self} as the retriever and DIN~\cite{zhou2018deep} as the ranker. 
% the pool size ($n$) of uniformly sampled items is set to 100, from which 20 items ($L$) are resampled for training retriever and ranker.
% In training, we firstly sampled itewhich 20 ms is set to 100, from y sample 100 items uniformly and then resample 20 items. 
In prediction, we first retrieve the top-100 items from all items using the retriever and then rank these items by the ranker for final outputs. For an extensive study of model agnostics, other choices for the retriever and ranker are also considered, as shown in Section~\ref{combinations}. The source code is released in github\footnote{https://github.com/AngusHuang17/CoRR\_www}. %All the experiments in the paper are conducted on RecStudio\footnote{https://github.com/ustcml/RecStudio} platform.

% \begin{multicols}{2}
%\vspace{-8pt}
\begin{figure}[hbt!]
    \centering
    \subfigure[Amazon(base:rand)]{
        % \vspace{-8pt}
        \includegraphics[width=0.48\columnwidth]{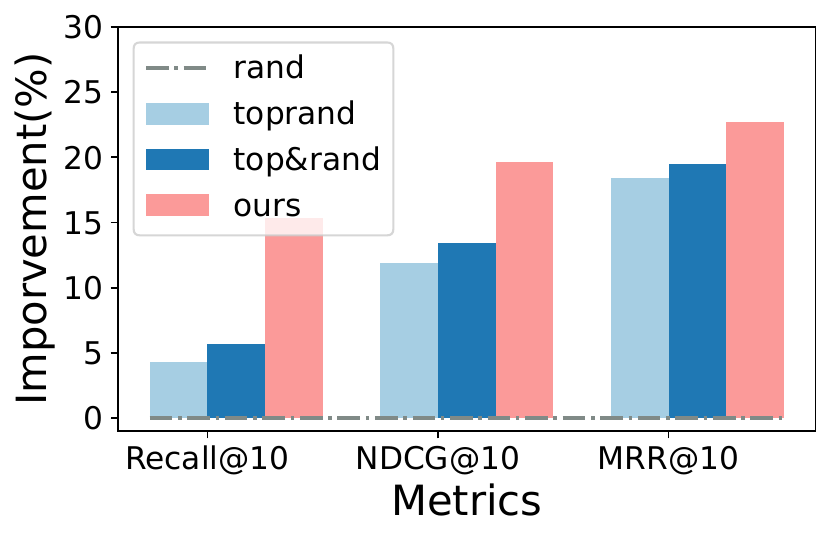}
    }
    \hspace{-0.3cm}
    \vspace{-8pt}
    \subfigure[Amazon(base:10 negatives)]{
        % \vspace{-8pt}
        \includegraphics[width=0.48\columnwidth]{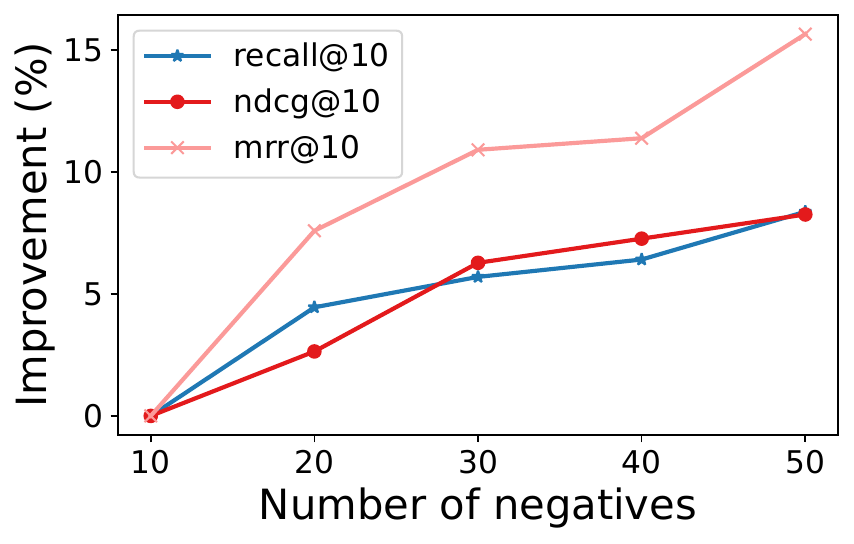}
    }
    \subfigure[Gowalla(base:rand)]{
        % \vspace{-8pt}
        \includegraphics[width=0.48\columnwidth]{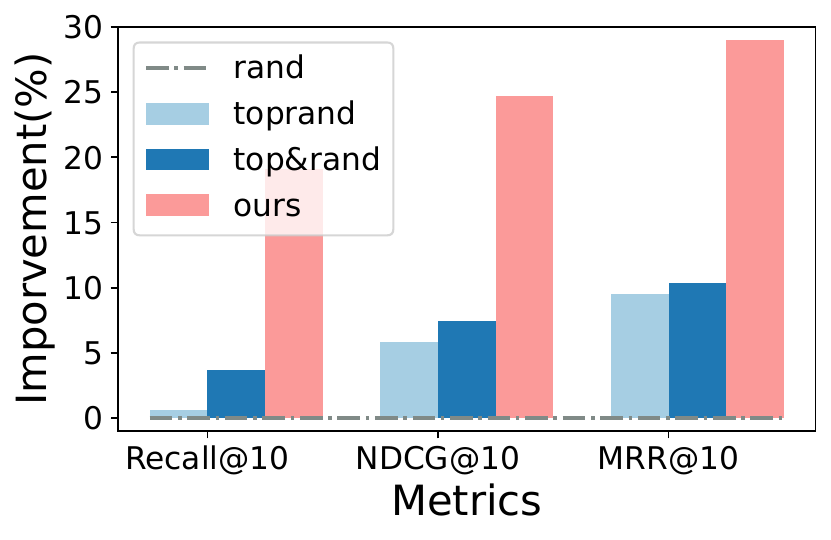}
    }
    \hspace{-0.3cm}
    \vspace{-8pt}
    \subfigure[Gowalla(base:10 negatives)]{
        % \vspace{-8pt}
        \includegraphics[width=0.48\columnwidth]{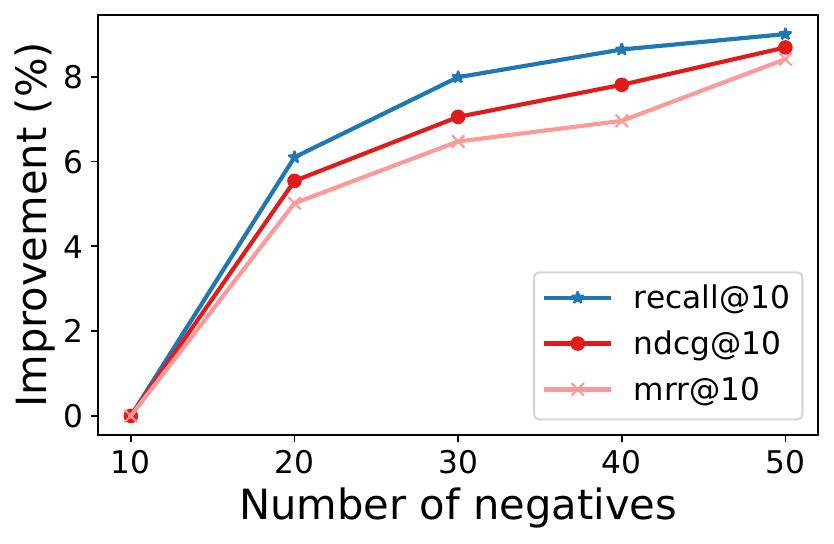}
    }
    \subfigure[MovieLens(base:rand)]{
        % \vspace{-8pt}
        \includegraphics[width=0.48\columnwidth]{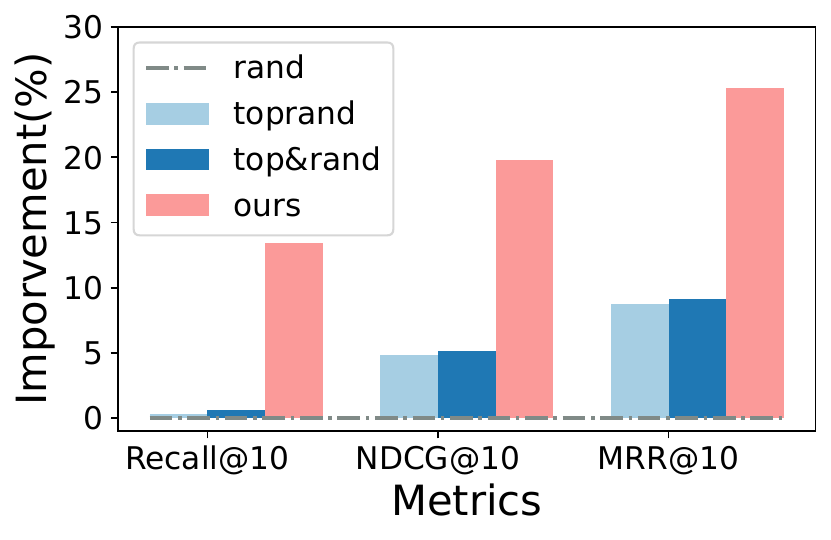}
    }
    \hspace{-0.3cm}
    \vspace{-8pt}
    \subfigure[MovieLens(base:10 negatives)]{
        % \vspace{-8pt}
        \includegraphics[width=0.48\columnwidth]{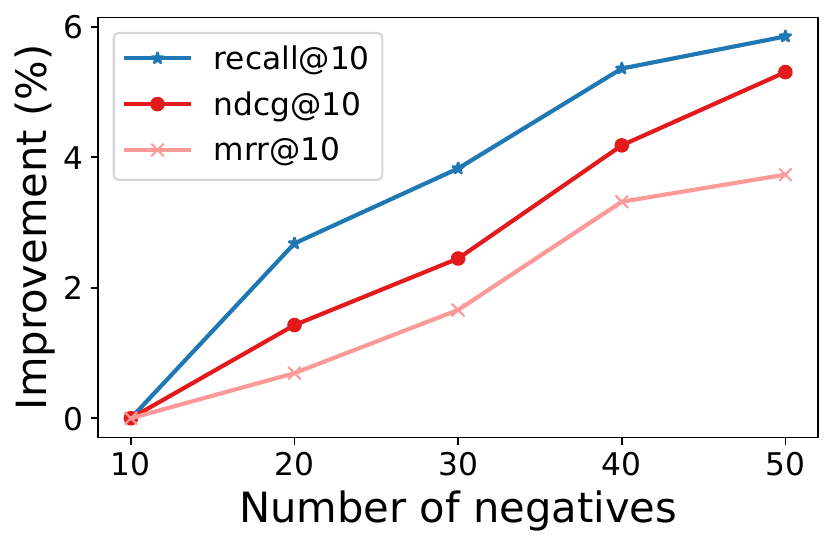}
    }
    \caption{Comparisons among different negative samples generating strategies (left column) and performance w.r.t negative items number (right column). The vertical axis represents the relative improvement to the base (\textit{rand} for the left column and \textit{10} for the right column).}
    \Description[Comparisons among different negative samples generating strategies (left column) and performance w.r.t negative items number (right column).]{The vertical axis represents the relative improvement to the base (rand for the left column and 10 for the right column). The left column indicates that ours negative samples generating strategy outperforms rand and top-k methods. And the right column indicates that the performance get better as the number of negative items increases.}
    \label{fig:sampling_neg}
    % \vspace{-0.1cm}
\end{figure}

\subsection{Comparison with Baselines}
\subsubsection{Baselines Methods}
We compare the proposed CoRR with four retriever methods and three existing training algorithms (i.e., ICC, RankFlow and Independent). Detailed information of baselines can be referred to in Appendix~\ref{sec:appd_baseline}.

\subsubsection{Results}
The experimental results on all datasets are reported in Table~\ref{tab1}. It shows that \textbf{CoRR outperforms both conventional DRS and the existing joint training methods with notable advantages}. The findings can be summarized as follows:

% \textit{Finding 1:} 
% \textcolor{red}{The ranker, i.e., DIN, shows significantly better performance than retrievers on all datasets}. Compared with the best retriever model, i.e., SASRec, DIN achieves an average relative 74.43\%, 67.79\% and 49.56\% improvement on all datasets in terms of NDCG@20, Recall@20 and MRR@20, respectively. The reason lies in that the ranker learns user preferences with the consideration of the relevance of user behaviors and the target item by deeper networks, which is more precise than indexable retrievers.
% Surprisingly, the ranker, i.e., DeepFM, DIN, shows slightly worse performance than retrievers on most datasets, which seems inconsistent with the assumption that the more expressive ranker would be more precise. We attribute the reason for the phenomena to that ranker is extremely good at rank on a small candidate item pool. To prove our conclusion, we conduct some experiments to compare the ability of retriever and ranker in two cases, for ranking all items and small candidate items. The result and settings are reported in \textcolor{red}{Section}.

\textit{Finding 1: The Independent training method performs better than all retrievers,} though both retriever and ranker are independently trained in The Independent method. In general recommendation, the Independent method outperforms the best retriever by {5.39\%, 5.16\%, 3.50\%} overall datasets in terms of Recall@10, NDCG@10 and MRR@10. Similarly in sequential recommendation, the improvements are up to {6.23\%, 4.11\%, 6.18\%}.
These results indicate that the limited expressiveness of the retriever can be compensated by the ranker and that the retriever and ranker can capture different information in the training data.
% Compared with the ranker, i.e., DIN, the Two-Stage method performs better on Amazon while having a close performance as DIN on other datasets. 

\textit{Finding 2: All three joint training methods (ICC, RankFlow and CoRR) almost outperform the Independent method.} The best of them -- CoRR's average relative improvements in terms of NDCG@10 to the Independent method are {25.62\%} and {18.57\%} in two tasks, respectively. This shows that modeling the collaboration between the retriever and ranker remarkably improves the cascading ranking systems. The better collaborative mechanism between them can lead to more improvements.  

\textit{Finding 3: Among three joint training methods, CoRR performs best in item recommendation.} This demonstrates the superiority of the proposed joint training framework. Compared with ICC, in which retriever and ranker are only learned with the training data, CoRR achieves {21.65\%} and {33.35\%} average relative improvements in terms of NDCG@10 on all datasets in general recommendation and sequential recommendation, respectively. Compared with RankFlow, in which retriever and ranker are jointly trained and reinforced by each other, CoRR still achieves {15.72\% and 11.62\%} average relative improvements in terms of NDCG@10 on all datasets in these two tasks. The results demonstrate that CoRR's cooperative training is more effective than RankFlow. More especially, the retriever can improve a lot by knowledge distillation from the ranker in CoRR while the retriever seldom gets improved in RankFlow.
% which is conducive for the ranker to learn better rankings among hard negatives.
% In addition, we conduct Students' t-test to demonstrate that CoRR significantly outperforms all other methods, detailed results are put in Appendix ~\ref{sec:appd_ttest}.

% \textit{Finding 4:} CoRR shows definitely better performance on the huge industrial dataset User Behavior, whose number of items reach more than one million.
% \textit{Finding 4:} CoRR shows better performance on datasets with larger numbers of items. On the Gowalla and Taobao dataset, whose numbers of items reach more than two hundred thousand, CoRR improves the performance by 26.21\% and 21.77\% in terms of NDCG@20, while there is only a relative 7.65\% and 1.75\% improvements on the Amazon and MovieLens-20M datasets compared with the adversarial strategy. It becomes more difficult to estimate the log-softmax with more items, but CoRR still has superior performance, indicating less bias of sampling to calculate the log softmax loss.

\subsection{Extensive Retrievers and Rankers}\label{combinations}
The proposed CoRR is a model-agnostic training framework, and we consider different combinations of retrievers and rankers to verify the universal effectiveness of CoRR. Experiments are conducted with four different retrievers (i.e., SASRec and Caser for sequential recommendation,  MF and DSSM for general recommendation) and four rankers (i.e., DIN and BST for sequential recommendation, DeepFM  and DCN for general recommendation) on the three datasets. We take the Independent and RankFlow as the baseline methods for comparison.
 
Table~\ref{tab6:extend} reports the results of the different combinations of retrievers and rankers. Our proposed CoRR, where the retriever and ranker are cooperatively trained, shows remarkably better performances than the Independent approach, leading to over 15\% improvements for any combinations of them. 
Compared with RankFlow, the relative improvements of CoRR can be up to 10\% for each any combinations of retrievers and rankers. 
This consistently validates the effectiveness of the proposed cooperative training framework and provides evidence of extensive practice for CoRR.

\subsection{Comparison of Different Negative Samplers} \label{sec:help_ranker}
The hard negative samples play important roles in the training of the rankers, and we compare different baseline strategies with the proposed sampling strategy in Section~\ref{sec:resample}, named \textit{Ours}. It is common to regard the top-k retrieval items from the retriever as hard negatives, so we compare our method with several top-k based strategies to verify the effectiveness of our method. They include 
\begin{itemize}[leftmargin=*]
    \item \textbf{Rand}: $n$ items are uniformly sampled from all items.
    \item \textbf{TopRand}: the top $5n$ items are retrieved by the retriever and then $n$ items are uniformly picked out from the retrieval items. 
    \item \textbf{Top\&Rand}: the top $n/2$ items are retrieved by the retriever and $n/2$ items are uniformly sampled from all items.
    \item \textbf{Ours}: sampling $n$ items from an adaptive and scalable sampler, by approximating softmax probability with subspace clustering. 
\end{itemize}
In the experiments, the number of negative samples $n$ is set to 20. Figure~\ref{fig:sampling_neg} reports the experimental results of these negative sampling strategies. It shows that \textbf{the adaptive sampler can generate higher-quality negative items than the exact top-k sampling method and variants, which remarkably improves model training}. The detailed findings are summarized as follows:

\textit{Finding 1: Both randomness and hardness are indispensable for sampling high-quality negative items.} The superiority of both \textit{TopRand} and \textit{Top\&Rand} to \textit{Rand} on all three datasets indicates the importance of hardness in negative sampling since the top-k retrieval items are harder and more informative than randomly sampled items. However, the top-k retrieval items are also highly probable to be false-negative, as evidenced by the superiority of \textit{Top\&Rand} to \textit{TopRand}, this indicates the importance of randomness in generating high-quality negative samples.
% \textit{Finding 1:} Introducing the randomly sampled items into the top-k items improves the performances. On one hand, the addition of randomly sampled items into the top-k items contributes to better performance, as evidenced by the fact that \textit{GTopRand} outperforms \textit{GTop} and \textit{LTopRand} outperforms \textit{LTop} on all three datasets. On the other hand, top-k items from the uniformly sampled set lead to superior recommendation quality than globally top-k items, since \textit{LTop}-based approaches show better performances than \textit{GTop}-based approaches. These experimental results substantially demonstrate that it is prone to encountering false negative samples from globally top-k items. These top-k items result in poor performance since they are static for training ranker in the lack of randomness. Moreover, the top-k items are sensitive to the retriever and may not be hard negatives in the case of a poor-performing retriever. 

\textit{Finding 2: The proposed adaptive sampler outperforms all negative sampling approaches.} This is evidenced by the superior recommendation performance of \textit{Ours} to other methods w.r.t all metrics on all datasets. On the one hand, the proposed sampler can ensure the randomness of sampled items by random sampling from the proposal distribution, such that any item can be sampled with significant probability as negative. In other words, the proposed sampling method can alleviate the false negative issue. On the other hand, the proposed sampler can sample the top-k retrieval items with higher probability than other items. This ensures the hardness of sampled items. Therefore, the recommendation model trained with these informative negative samples can converge to a better solution, generating higher-quality recommendation results. 

\begin{table}[!htb]
    % \vspace{-0.1cm}
    \caption{Performance w.r.t. KL loss. ($\times 10^{-2}$)}
    \setlength\tabcolsep{3pt}
    % \usepackage{multirow}
    \renewcommand\arraystretch{0.9}
    % \resizebox{\textwidth}{!}{
    \begin{tabular}{l|l|cc|cc}
    \toprule
    \multirow{2}{*}{Dataset} & \multirow{2}{*}{Metric} & \multicolumn{2}{c|}{w/o KL} & \multicolumn{2}{c}{KL} \\ \cmidrule{3-6} 
                             & & Retriever     & CoRR     & Retriever   & CoRR  \\ \midrule
    \multirow{3}{*}{Amazon} & R@10 & 4.88 & 4.96 & 5.25 & 5.87 \\
           & N@10 & 2.38 & 2.56 & 2.63 & 3.11 \\ 
           & M@10 & 1.84 & 2.14 & 1.90 & 2.27 \\ \midrule
    \multirow{3}{*}{Gowalla} & R@10 & 13.10 & 13.42 & 13.90 & 14.61   \\
            & N@10 & 6.65 & 6.97 & 7.88 & 8.38   \\ 
            & M@10 & 5.69 & 6.00 & 6.05 & 6.49   \\ \midrule
    \multirow{3}{*}{MovieLens}  & R@10 & 15.96 & 17.80 & 16.66 & 18.77   \\
               & N@10 & 7.63 & 8.96 & 7.82 & 9.94   \\ 
               & M@10 & 5.69 & 6.60 & 6.31 & 7.28   \\  \bottomrule
    \end{tabular}
    % }
    \label{tab:kl_loss}
    % \vspace{-0.1cm}
\end{table}

\subsection{Effect of Knowledge Distillation} \label{sec:help_retriever}
In order to verify the effect of the distillation loss (Equation~\ref{eq:kl_loss}) in Section~\ref{sec:distillation}, we compare CoRR with its variant by removing distillation loss when training the retriever (w/o KL in Table~\ref{tab:kl_loss}). We consider the recommendation performance of the retriever (the Retriever column) and the two-stage framework (the CoRR column) in this case. The results are shown in Table~\ref{tab:kl_loss}.

The results show that \textbf{the ranking-order preserving distillation loss indeed remarkably improve the retriever}. The use of the sampled KL divergence can contribute to 10.89\%, 17.55\%, 8.18\% average relative improvements w.r.t NDCG@10, Recall@10 and MRR@10. According to the superior performance of CoRR to RankFlow in Table~\ref{tab1}, the ranking-order preserving distillation in CoRR improves the retriever, while tutor learning in RankFlow seldom takes effect. This demonstrates the effectiveness of the proposed sampled KL divergence for knowledge distillation.

% As shown in Table~\ref{tab4}, the proposed \textit{Resample} outperforms the top-k based approaches and randomly sampling approaches. \textit{Resample} achieves a relative 0.8\%, 0.4\% improvements in terms of NDCG@20 and MRR@20, respectively, on all three datasets. This corresponds with the theoretical analysis in Theorem\ref{thm:kl_convergence}. The asymptotic-unbiased approximation of KL divergence compensates the bias resulting from the sampled items, leading to better convergence for training the retriever. Compared with previous cooperative training frameworks, such as Adversarial and Two-Stage, the retriever also receives signals from rankers, instead of being trained standalone, providing better hard negative samples as the model improves.

% \begin{figure}[ht]
%     \centering
%     \subfigure[Amazon]{
%         \includegraphics[width=0.48\columnwidth]{./fig/num_neg_Amazon.pdf}
%     }
%     \hspace{-0.3cm}
%     \subfigure[Gowalla]{
%         \includegraphics[width=0.48\columnwidth]{./fig/num_neg_Gowalla.pdf}
%     }
%     \subfigure[MovieLens]{
%         \includegraphics[width=0.48\columnwidth]{./fig/num_neg_MovieLens.pdf}
% }
%     \vspace{-0.4cm}
%     \caption{Performance w.r.t. negative items number}
%     \label{fig:neg_num}
% \end{figure}

\subsection{Sensitivity to the Number of Negatives}
As discussed in Section~\ref{sec:resample} and ~\ref{sec:distillation}, we provide the theoretical results of asymptotic-unbiased estimation of the KL divergence and the log-softmax. To further investigate the influence of the negative number, we conduct experiments on the three datasets where the negative number is varied in $\{10, 20, 30, 40, 50\}$. The results are shown in the right column of Figure~\ref{fig:sampling_neg}.

These figures show that CoRR has consistently better performance with the increasing number of negative samples, being in line with the theoretical results detailed in Section~\ref{sec:distillation}. When more items are sampled for approximating the KL divergence and log-softmax loss, their estimation bias gets smaller, so that the both retrievers and rankers are better trained.

\section{Conclusion}
In this paper, we propose a novel DRS joint training framework CoRR, where the retriever and ranker are made mutually reinforced. We develop an adaptive and scalable sampler based on the retriever, which generates hard negative samples to facilitate the ranker's training. We also propose a novel asymptotic-unbiased estimation of KL divergence, which improves the effect of knowledge distillation, and thus contributes to the retriever's trainingÍ. Comprehensive experiments over four large-scale datasets verify the effectiveness of CoRR, as it outperforms both conventional DRS and the existing joint training methods with notable advantages.

\begin{acks}
The work was supported by grants from the National Key R\&D Program of China under Grant No. 2020AAA0103800, the National Natural Science Foundation of China (No. 62022077).
\end{acks}

\bibliographystyle{ACM-Reference-Format}
\bibliography{ref}


\begin{thebibliography}{53}

%%% the Web address.

\ifx \showCODEN    \undefined \def \showCODEN     #1{\unskip}     \fi
\ifx \showDOI      \undefined \def \showDOI       #1{#1}\fi
\ifx \showISBNx    \undefined \def \showISBNx     #1{\unskip}     \fi
\ifx \showISBNxiii \undefined \def \showISBNxiii  #1{\unskip}     \fi
\ifx \showISSN     \undefined \def \showISSN      #1{\unskip}     \fi
\ifx \showLCCN     \undefined \def \showLCCN      #1{\unskip}     \fi
\ifx \shownote     \undefined \def \shownote      #1{#1}          \fi
\ifx \showarticletitle \undefined \def \showarticletitle #1{#1}   \fi
\ifx \showURL      \undefined \def \showURL       {\relax}        \fi
% invisible to TeX
\providecommand\bibfield[2]{#2}
\providecommand\bibinfo[2]{#2}
\providecommand\natexlab[1]{#1}
\providecommand\showeprint[2][]{arXiv:#2}

\bibitem[Bengio and Sen{\'e}cal(2008)]%
        {bengio2008adaptive}
\bibfield{author}{\bibinfo{person}{Yoshua Bengio} {and}
  \bibinfo{person}{Jean-S{\'e}bastien Sen{\'e}cal}.}
  \bibinfo{year}{2008}\natexlab{}.
\newblock \showarticletitle{Adaptive importance sampling to accelerate training
  of a neural probabilistic language model}.
\newblock \bibinfo{journal}{\emph{IEEE Transactions on Neural Networks}}
  \bibinfo{volume}{19}, \bibinfo{number}{4} (\bibinfo{year}{2008}),
  \bibinfo{pages}{713--722}.
\newblock


\bibitem[Blanc and Rendle(2018)]%
        {blanc2018adaptive}
\bibfield{author}{\bibinfo{person}{Guy Blanc} {and} \bibinfo{person}{Steffen
  Rendle}.} \bibinfo{year}{2018}\natexlab{}.
\newblock \showarticletitle{Adaptive sampled softmax with kernel based
  sampling}. In \bibinfo{booktitle}{\emph{International Conference on Machine
  Learning}}. PMLR, \bibinfo{pages}{590--599}.
\newblock


\bibitem[Bruch et~al\mbox{.}(2019)]%
        {bruch2019analysis}
\bibfield{author}{\bibinfo{person}{Sebastian Bruch}, \bibinfo{person}{Xuanhui
  Wang}, \bibinfo{person}{Michael Bendersky}, {and} \bibinfo{person}{Marc
  Najork}.} \bibinfo{year}{2019}\natexlab{}.
\newblock \showarticletitle{An analysis of the softmax cross entropy loss for
  learning-to-rank with binary relevance}. In
  \bibinfo{booktitle}{\emph{Proceedings of the 2019 ACM SIGIR international
  conference on theory of information retrieval}}. \bibinfo{pages}{75--78}.
\newblock


\bibitem[Cen et~al\mbox{.}(2020)]%
        {cen2020controllable}
\bibfield{author}{\bibinfo{person}{Yukuo Cen}, \bibinfo{person}{Jianwei Zhang},
  \bibinfo{person}{Xu Zou}, \bibinfo{person}{Chang Zhou},
  \bibinfo{person}{Hongxia Yang}, {and} \bibinfo{person}{Jie Tang}.}
  \bibinfo{year}{2020}\natexlab{}.
\newblock \showarticletitle{Controllable multi-interest framework for
  recommendation}. In \bibinfo{booktitle}{\emph{Proceedings of the 26th ACM
  SIGKDD International Conference on Knowledge Discovery \& Data Mining}}.
  \bibinfo{pages}{2942--2951}.
\newblock


\bibitem[Chen et~al\mbox{.}(2021)]%
        {chen2021fast}
\bibfield{author}{\bibinfo{person}{Jin Chen}, \bibinfo{person}{Binbin Jin},
  \bibinfo{person}{Xu Huang}, \bibinfo{person}{Defu Lian}, \bibinfo{person}{Kai
  Zheng}, {and} \bibinfo{person}{Enhong Chen}.}
  \bibinfo{year}{2021}\natexlab{}.
\newblock \showarticletitle{Fast Variational AutoEncoder with Inverted
  Multi-Index for Collaborative Filtering}.
\newblock \bibinfo{journal}{\emph{arXiv preprint arXiv:2109.05773}}
  (\bibinfo{year}{2021}).
\newblock


\bibitem[Chen et~al\mbox{.}(2017)]%
        {chen2017efficient}
\bibfield{author}{\bibinfo{person}{Ruey-Cheng Chen}, \bibinfo{person}{Luke
  Gallagher}, \bibinfo{person}{Roi Blanco}, {and} \bibinfo{person}{J~Shane
  Culpepper}.} \bibinfo{year}{2017}\natexlab{}.
\newblock \showarticletitle{Efficient cost-aware cascade ranking in multi-stage
  retrieval}. In \bibinfo{booktitle}{\emph{Proceedings of the 40th
  International ACM SIGIR Conference on Research and Development in Information
  Retrieval}}. \bibinfo{pages}{445--454}.
\newblock


\bibitem[Covington et~al\mbox{.}(2016)]%
        {covington2016deep}
\bibfield{author}{\bibinfo{person}{Paul Covington}, \bibinfo{person}{Jay
  Adams}, {and} \bibinfo{person}{Emre Sargin}.}
  \bibinfo{year}{2016}\natexlab{}.
\newblock \showarticletitle{Deep neural networks for youtube recommendations}.
  In \bibinfo{booktitle}{\emph{Proceedings of the 10th ACM conference on
  recommender systems}}. \bibinfo{pages}{191--198}.
\newblock


\bibitem[Fan et~al\mbox{.}(2019)]%
        {fan2019mobius}
\bibfield{author}{\bibinfo{person}{Miao Fan}, \bibinfo{person}{Jiacheng Guo},
  \bibinfo{person}{Shuai Zhu}, \bibinfo{person}{Shuo Miao},
  \bibinfo{person}{Mingming Sun}, {and} \bibinfo{person}{Ping Li}.}
  \bibinfo{year}{2019}\natexlab{}.
\newblock \showarticletitle{MOBIUS: towards the next generation of query-ad
  matching in baidu's sponsored search}. In
  \bibinfo{booktitle}{\emph{Proceedings of the 25th ACM SIGKDD International
  Conference on Knowledge Discovery \& Data Mining}}.
  \bibinfo{pages}{2509--2517}.
\newblock


\bibitem[Feng et~al\mbox{.}(2022)]%
        {feng2022recommender}
\bibfield{author}{\bibinfo{person}{Chao Feng}, \bibinfo{person}{Wuchao Li},
  \bibinfo{person}{Defu Lian}, \bibinfo{person}{Zheng Liu}, {and}
  \bibinfo{person}{Enhong Chen}.} \bibinfo{year}{2022}\natexlab{}.
\newblock \showarticletitle{Recommender Forest for Efficient Retrieval}. In
  \bibinfo{booktitle}{\emph{Advances in Neural Information Processing
  Systems}}, \bibfield{editor}{\bibinfo{person}{Alice~H. Oh},
  \bibinfo{person}{Alekh Agarwal}, \bibinfo{person}{Danielle Belgrave}, {and}
  \bibinfo{person}{Kyunghyun Cho}} (Eds.).
\newblock
\urldef\tempurl%
\url{https://openreview.net/forum?id=Yc4MjP2Mnob}
\showURL{%
\tempurl}


\bibitem[Feng et~al\mbox{.}(2023)]%
        {feng2023reinforcement}
\bibfield{author}{\bibinfo{person}{Chao Feng}, \bibinfo{person}{Defu Lian},
  \bibinfo{person}{Xiting Wang}, \bibinfo{person}{Zheng Liu},
  \bibinfo{person}{Xing Xie}, {and} \bibinfo{person}{Enhong Chen}.}
  \bibinfo{year}{2023}\natexlab{}.
\newblock \showarticletitle{Reinforcement Routing on Proximity Graph for
  Efficient Recommendation}.
\newblock  \bibinfo{volume}{41}, \bibinfo{number}{1}, Article
  \bibinfo{articleno}{8} (\bibinfo{date}{jan} \bibinfo{year}{2023}),
  \bibinfo{numpages}{27}~pages.
\newblock
\showISSN{1046-8188}
\urldef\tempurl%
\url{https://doi.org/10.1145/3512767}
\showDOI{\tempurl}


\bibitem[Gallagher et~al\mbox{.}(2019)]%
        {gallagher2019joint}
\bibfield{author}{\bibinfo{person}{Luke Gallagher}, \bibinfo{person}{Ruey-Cheng
  Chen}, \bibinfo{person}{Roi Blanco}, {and} \bibinfo{person}{J.~Shane
  Culpepper}.} \bibinfo{year}{2019}\natexlab{}.
\newblock \showarticletitle{Joint Optimization of Cascade Ranking Models}. In
  \bibinfo{booktitle}{\emph{Proceedings of the Twelfth ACM International
  Conference on Web Search and Data Mining}} (Melbourne VIC, Australia)
  \emph{(\bibinfo{series}{WSDM '19})}. \bibinfo{publisher}{Association for
  Computing Machinery}, \bibinfo{address}{New York, NY, USA},
  \bibinfo{pages}{15–23}.
\newblock
\showISBNx{9781450359405}
\urldef\tempurl%
\url{https://doi.org/10.1145/3289600.3290986}
\showDOI{\tempurl}


\bibitem[Guo et~al\mbox{.}(2017)]%
        {guo2017deepfm}
\bibfield{author}{\bibinfo{person}{Huifeng Guo}, \bibinfo{person}{Ruiming
  Tang}, \bibinfo{person}{Yunming Ye}, \bibinfo{person}{Zhenguo Li}, {and}
  \bibinfo{person}{Xiuqiang He}.} \bibinfo{year}{2017}\natexlab{}.
\newblock \showarticletitle{DeepFM: a factorization-machine based neural
  network for CTR prediction}. In \bibinfo{booktitle}{\emph{Proceedings of
  IJCAI'17}}. AAAI Press, \bibinfo{pages}{1725--1731}.
\newblock


\bibitem[Guo et~al\mbox{.}(2020)]%
        {guo2020accelerating}
\bibfield{author}{\bibinfo{person}{Ruiqi Guo}, \bibinfo{person}{Philip Sun},
  \bibinfo{person}{Erik Lindgren}, \bibinfo{person}{Quan Geng},
  \bibinfo{person}{David Simcha}, \bibinfo{person}{Felix Chern}, {and}
  \bibinfo{person}{Sanjiv Kumar}.} \bibinfo{year}{2020}\natexlab{}.
\newblock \showarticletitle{Accelerating large-scale inference with anisotropic
  vector quantization}. In \bibinfo{booktitle}{\emph{International Conference
  on Machine Learning}}. PMLR, \bibinfo{pages}{3887--3896}.
\newblock


\bibitem[He et~al\mbox{.}(2017)]%
        {he2017neural}
\bibfield{author}{\bibinfo{person}{Xiangnan He}, \bibinfo{person}{Lizi Liao},
  \bibinfo{person}{Hanwang Zhang}, \bibinfo{person}{Liqiang Nie},
  \bibinfo{person}{Xia Hu}, {and} \bibinfo{person}{Tat-Seng Chua}.}
  \bibinfo{year}{2017}\natexlab{}.
\newblock \showarticletitle{Neural collaborative filtering}. In
  \bibinfo{booktitle}{\emph{Proceedings of the 26th international conference on
  world wide web}}. \bibinfo{pages}{173--182}.
\newblock


\bibitem[Hidasi et~al\mbox{.}(2015)]%
        {hidasi2015session}
\bibfield{author}{\bibinfo{person}{Bal{\'a}zs Hidasi},
  \bibinfo{person}{Alexandros Karatzoglou}, \bibinfo{person}{Linas Baltrunas},
  {and} \bibinfo{person}{Domonkos Tikk}.} \bibinfo{year}{2015}\natexlab{}.
\newblock \showarticletitle{Session-based recommendations with recurrent neural
  networks}.
\newblock \bibinfo{journal}{\emph{arXiv preprint arXiv:1511.06939}}
  (\bibinfo{year}{2015}).
\newblock


\bibitem[Hinton et~al\mbox{.}(2015)]%
        {hinton2015distilling}
\bibfield{author}{\bibinfo{person}{Geoffrey Hinton}, \bibinfo{person}{Oriol
  Vinyals}, \bibinfo{person}{Jeff Dean}, {et~al\mbox{.}}}
  \bibinfo{year}{2015}\natexlab{}.
\newblock \showarticletitle{Distilling the knowledge in a neural network}.
\newblock \bibinfo{journal}{\emph{arXiv preprint arXiv:1503.02531}}
  \bibinfo{volume}{2}, \bibinfo{number}{7} (\bibinfo{year}{2015}).
\newblock


\bibitem[Hron et~al\mbox{.}(2021)]%
        {hron2021component}
\bibfield{author}{\bibinfo{person}{Jiri Hron}, \bibinfo{person}{Karl Krauth},
  \bibinfo{person}{Michael Jordan}, {and} \bibinfo{person}{Niki Kilbertus}.}
  \bibinfo{year}{2021}\natexlab{}.
\newblock \showarticletitle{On component interactions in two-stage recommender
  systems}.
\newblock \bibinfo{journal}{\emph{Advances in Neural Information Processing
  Systems}}  \bibinfo{volume}{34} (\bibinfo{year}{2021}),
  \bibinfo{pages}{2744--2757}.
\newblock


\bibitem[Hsieh et~al\mbox{.}(2017)]%
        {hsieh2017colla}
\bibfield{author}{\bibinfo{person}{Cheng-Kang Hsieh}, \bibinfo{person}{Longqi
  Yang}, \bibinfo{person}{Yin Cui}, \bibinfo{person}{Tsung-Yi Lin},
  \bibinfo{person}{Serge Belongie}, {and} \bibinfo{person}{Deborah Estrin}.}
  \bibinfo{year}{2017}\natexlab{}.
\newblock \showarticletitle{Collaborative metric learning}. In
  \bibinfo{booktitle}{\emph{Proceedings of the 26th international conference on
  world wide web}}. \bibinfo{pages}{193--201}.
\newblock


\bibitem[Hu et~al\mbox{.}(2008)]%
        {hu2008collaborative}
\bibfield{author}{\bibinfo{person}{Y. Hu}, \bibinfo{person}{Y. Koren}, {and}
  \bibinfo{person}{C. Volinsky}.} \bibinfo{year}{2008}\natexlab{}.
\newblock \showarticletitle{Collaborative filtering for implicit feedback
  datasets}. In \bibinfo{booktitle}{\emph{Proceedings of ICDM'08}}. IEEE,
  \bibinfo{pages}{263--272}.
\newblock


\bibitem[Huang et~al\mbox{.}(2013)]%
        {huang2013learning}
\bibfield{author}{\bibinfo{person}{Po-Sen Huang}, \bibinfo{person}{Xiaodong
  He}, \bibinfo{person}{Jianfeng Gao}, \bibinfo{person}{Li Deng},
  \bibinfo{person}{Alex Acero}, {and} \bibinfo{person}{Larry Heck}.}
  \bibinfo{year}{2013}\natexlab{}.
\newblock \showarticletitle{Learning deep structured semantic models for web
  search using clickthrough data}. In \bibinfo{booktitle}{\emph{Proceedings of
  CIKM'13}}. ACM, \bibinfo{pages}{2333--2338}.
\newblock


\bibitem[Jin et~al\mbox{.}(2020)]%
        {jin2020sampling}
\bibfield{author}{\bibinfo{person}{Binbin Jin}, \bibinfo{person}{Defu Lian},
  \bibinfo{person}{Zheng Liu}, \bibinfo{person}{Qi Liu},
  \bibinfo{person}{Jianhui Ma}, \bibinfo{person}{Xing Xie}, {and}
  \bibinfo{person}{Enhong Chen}.} \bibinfo{year}{2020}\natexlab{}.
\newblock \bibinfo{title}{Sampling-Decomposable Generative Adversarial
  Recommender}.
\newblock
\newblock
\urldef\tempurl%
\url{https://doi.org/10.48550/ARXIV.2011.00956}
\showDOI{\tempurl}


\bibitem[Johnson(2014)]%
        {Johnson2014LogisticMF}
\bibfield{author}{\bibinfo{person}{Christopher~C. Johnson}.}
  \bibinfo{year}{2014}\natexlab{}.
\newblock \showarticletitle{Logistic Matrix Factorization for Implicit Feedback
  Data}.
\newblock


\bibitem[Johnson et~al\mbox{.}(2017)]%
        {JDH17}
\bibfield{author}{\bibinfo{person}{Jeff Johnson}, \bibinfo{person}{Matthijs
  Douze}, {and} \bibinfo{person}{Herv{\'e} J{\'e}gou}.}
  \bibinfo{year}{2017}\natexlab{}.
\newblock \showarticletitle{Billion-scale similarity search with GPUs}.
\newblock \bibinfo{journal}{\emph{arXiv preprint arXiv:1702.08734}}
  (\bibinfo{year}{2017}).
\newblock


\bibitem[Kang et~al\mbox{.}(2020)]%
        {kang2020rrd}
\bibfield{author}{\bibinfo{person}{SeongKu Kang}, \bibinfo{person}{Junyoung
  Hwang}, \bibinfo{person}{Wonbin Kweon}, {and} \bibinfo{person}{Hwanjo Yu}.}
  \bibinfo{year}{2020}\natexlab{}.
\newblock \showarticletitle{DE-RRD: A knowledge distillation framework for
  recommender system}. In \bibinfo{booktitle}{\emph{Proceedings of the 29th ACM
  International Conference on Information \& Knowledge Management}}.
  \bibinfo{pages}{605--614}.
\newblock


\bibitem[Kang et~al\mbox{.}(2021)]%
        {kang2021topology}
\bibfield{author}{\bibinfo{person}{SeongKu Kang}, \bibinfo{person}{Junyoung
  Hwang}, \bibinfo{person}{Wonbin Kweon}, {and} \bibinfo{person}{Hwanjo Yu}.}
  \bibinfo{year}{2021}\natexlab{}.
\newblock \showarticletitle{Topology distillation for recommender system}. In
  \bibinfo{booktitle}{\emph{Proceedings of the 27th ACM SIGKDD Conference on
  Knowledge Discovery \& Data Mining}}. \bibinfo{pages}{829--839}.
\newblock


\bibitem[Kang and McAuley(2018)]%
        {kang2018self}
\bibfield{author}{\bibinfo{person}{Wang-Cheng Kang} {and}
  \bibinfo{person}{Julian McAuley}.} \bibinfo{year}{2018}\natexlab{}.
\newblock \showarticletitle{Self-attentive sequential recommendation}. In
  \bibinfo{booktitle}{\emph{2018 IEEE International Conference on Data Mining
  (ICDM)}}. IEEE, \bibinfo{pages}{197--206}.
\newblock


\bibitem[Koren et~al\mbox{.}(2009)]%
        {5197422}
\bibfield{author}{\bibinfo{person}{Yehuda Koren}, \bibinfo{person}{Robert
  Bell}, {and} \bibinfo{person}{Chris Volinsky}.}
  \bibinfo{year}{2009}\natexlab{}.
\newblock \showarticletitle{Matrix Factorization Techniques for Recommender
  Systems}.
\newblock \bibinfo{journal}{\emph{Computer}} \bibinfo{volume}{42},
  \bibinfo{number}{8} (\bibinfo{year}{2009}), \bibinfo{pages}{30--37}.
\newblock
\urldef\tempurl%
\url{https://doi.org/10.1109/MC.2009.263}
\showDOI{\tempurl}


\bibitem[Kweon et~al\mbox{.}(2021)]%
        {kweon2021bidirectional}
\bibfield{author}{\bibinfo{person}{Wonbin Kweon}, \bibinfo{person}{SeongKu
  Kang}, {and} \bibinfo{person}{Hwanjo Yu}.} \bibinfo{year}{2021}\natexlab{}.
\newblock \showarticletitle{Bidirectional Distillation for Top-K Recommender
  System}. In \bibinfo{booktitle}{\emph{Proceedings of the Web Conference
  2021}}. \bibinfo{pages}{3861--3871}.
\newblock


\bibitem[Lee et~al\mbox{.}(2019)]%
        {lee2019collaborative}
\bibfield{author}{\bibinfo{person}{Jae-woong Lee}, \bibinfo{person}{Minjin
  Choi}, \bibinfo{person}{Jongwuk Lee}, {and} \bibinfo{person}{Hyunjung Shim}.}
  \bibinfo{year}{2019}\natexlab{}.
\newblock \showarticletitle{Collaborative distillation for top-N
  recommendation}. In \bibinfo{booktitle}{\emph{2019 IEEE International
  Conference on Data Mining (ICDM)}}. IEEE, \bibinfo{pages}{369--378}.
\newblock


\bibitem[Lian et~al\mbox{.}(2020a)]%
        {lian2020personalized}
\bibfield{author}{\bibinfo{person}{Defu Lian}, \bibinfo{person}{Qi Liu}, {and}
  \bibinfo{person}{Enhong Chen}.} \bibinfo{year}{2020}\natexlab{a}.
\newblock \showarticletitle{Personalized Ranking with Importance Sampling}. In
  \bibinfo{booktitle}{\emph{Proceedings of The Web Conference 2020}} (Taipei,
  Taiwan) \emph{(\bibinfo{series}{WWW '20})}. \bibinfo{publisher}{Association
  for Computing Machinery}, \bibinfo{address}{New York, NY, USA},
  \bibinfo{pages}{1093–1103}.
\newblock
\showISBNx{9781450370233}
\urldef\tempurl%
\url{https://doi.org/10.1145/3366423.3380187}
\showDOI{\tempurl}


\bibitem[Lian et~al\mbox{.}(2020b)]%
        {lian2020lightrec}
\bibfield{author}{\bibinfo{person}{Defu Lian}, \bibinfo{person}{Haoyu Wang},
  \bibinfo{person}{Zheng Liu}, \bibinfo{person}{Jianxun Lian},
  \bibinfo{person}{Enhong Chen}, {and} \bibinfo{person}{Xing Xie}.}
  \bibinfo{year}{2020}\natexlab{b}.
\newblock \showarticletitle{LightRec: A Memory and Search-Efficient Recommender
  System}. In \bibinfo{booktitle}{\emph{Proceedings of The Web Conference
  2020}} (Taipei, Taiwan) \emph{(\bibinfo{series}{WWW '20})}.
  \bibinfo{publisher}{Association for Computing Machinery},
  \bibinfo{address}{New York, NY, USA}, \bibinfo{pages}{695–705}.
\newblock
\showISBNx{9781450370233}
\urldef\tempurl%
\url{https://doi.org/10.1145/3366423.3380151}
\showDOI{\tempurl}


\bibitem[Lian et~al\mbox{.}(2020c)]%
        {lian2020geography}
\bibfield{author}{\bibinfo{person}{Defu Lian}, \bibinfo{person}{Yongji Wu},
  \bibinfo{person}{Yong Ge}, \bibinfo{person}{Xing Xie}, {and}
  \bibinfo{person}{Enhong Chen}.} \bibinfo{year}{2020}\natexlab{c}.
\newblock \showarticletitle{Geography-Aware Sequential Location Recommendation}
  \emph{(\bibinfo{series}{KDD '20})}. \bibinfo{publisher}{Association for
  Computing Machinery}, \bibinfo{address}{New York, NY, USA},
  \bibinfo{pages}{2009–2019}.
\newblock
\showISBNx{9781450379984}
\urldef\tempurl%
\url{https://doi.org/10.1145/3394486.3403252}
\showDOI{\tempurl}


\bibitem[Liang et~al\mbox{.}(2018)]%
        {liang2018variational}
\bibfield{author}{\bibinfo{person}{Dawen Liang}, \bibinfo{person}{Rahul~G
  Krishnan}, \bibinfo{person}{Matthew~D Hoffman}, {and} \bibinfo{person}{Tony
  Jebara}.} \bibinfo{year}{2018}\natexlab{}.
\newblock \showarticletitle{Variational Autoencoders for Collaborative
  Filtering}. In \bibinfo{booktitle}{\emph{Proceedings of WWW'18}}.
  International World Wide Web Conferences Steering Committee,
  \bibinfo{pages}{689--698}.
\newblock


\bibitem[Qin et~al\mbox{.}(2022)]%
        {qin2022rankflow}
\bibfield{author}{\bibinfo{person}{Jiarui Qin}, \bibinfo{person}{Jiachen Zhu},
  \bibinfo{person}{Bo Chen}, \bibinfo{person}{Zhirong Liu},
  \bibinfo{person}{Weiwen Liu}, \bibinfo{person}{Ruiming Tang},
  \bibinfo{person}{Rui Zhang}, \bibinfo{person}{Yong Yu}, {and}
  \bibinfo{person}{Weinan Zhang}.} \bibinfo{year}{2022}\natexlab{}.
\newblock \showarticletitle{RankFlow: Joint Optimization of Multi-Stage Cascade
  Ranking Systems as Flows}. In \bibinfo{booktitle}{\emph{Proceedings of the
  45th International ACM SIGIR Conference on Research and Development in
  Information Retrieval}} (Madrid, Spain) \emph{(\bibinfo{series}{SIGIR '22})}.
  \bibinfo{publisher}{Association for Computing Machinery},
  \bibinfo{address}{New York, NY, USA}, \bibinfo{pages}{814–824}.
\newblock
\showISBNx{9781450387323}
\urldef\tempurl%
\url{https://doi.org/10.1145/3477495.3532050}
\showDOI{\tempurl}


\bibitem[Rendle and Freudenthaler(2014)]%
        {rendle2014improving}
\bibfield{author}{\bibinfo{person}{Steffen Rendle} {and}
  \bibinfo{person}{Christoph Freudenthaler}.} \bibinfo{year}{2014}\natexlab{}.
\newblock \showarticletitle{Improving pairwise learning for item recommendation
  from implicit feedback}. In \bibinfo{booktitle}{\emph{Proceedings of the 7th
  ACM international conference on Web search and data mining}}.
  \bibinfo{pages}{273--282}.
\newblock


\bibitem[Rendle et~al\mbox{.}(2009)]%
        {rendle2009bpr}
\bibfield{author}{\bibinfo{person}{S. Rendle}, \bibinfo{person}{C.
  Freudenthaler}, \bibinfo{person}{Z. Gantner}, {and} \bibinfo{person}{L.
  Schmidt-Thieme}.} \bibinfo{year}{2009}\natexlab{}.
\newblock \showarticletitle{BPR: Bayesian personalized ranking from implicit
  feedback}. In \bibinfo{booktitle}{\emph{Proceedings of UAI'09}}. AUAI Press,
  \bibinfo{pages}{452--461}.
\newblock


\bibitem[Spring and Shrivastava(2017)]%
        {spring2017new}
\bibfield{author}{\bibinfo{person}{Ryan Spring} {and}
  \bibinfo{person}{Anshumali Shrivastava}.} \bibinfo{year}{2017}\natexlab{}.
\newblock \showarticletitle{A new unbiased and efficient class of lsh-based
  samplers and estimators for partition function computation in log-linear
  models}.
\newblock \bibinfo{journal}{\emph{arXiv preprint arXiv:1703.05160}}
  (\bibinfo{year}{2017}).
\newblock


\bibitem[Sun et~al\mbox{.}(2019b)]%
        {sun2019bert4rec}
\bibfield{author}{\bibinfo{person}{Fei Sun}, \bibinfo{person}{Jun Liu},
  \bibinfo{person}{Jian Wu}, \bibinfo{person}{Changhua Pei},
  \bibinfo{person}{Xiao Lin}, \bibinfo{person}{Wenwu Ou}, {and}
  \bibinfo{person}{Peng Jiang}.} \bibinfo{year}{2019}\natexlab{b}.
\newblock \showarticletitle{BERT4Rec: Sequential Recommendation with
  Bidirectional Encoder Representations from Transformer}. In
  \bibinfo{booktitle}{\emph{Proceedings of the 28th ACM International
  Conference on Information and Knowledge Management}} (Beijing, China)
  \emph{(\bibinfo{series}{CIKM '19})}. \bibinfo{publisher}{Association for
  Computing Machinery}, \bibinfo{address}{New York, NY, USA},
  \bibinfo{pages}{1441–1450}.
\newblock
\showISBNx{9781450369763}
\urldef\tempurl%
\url{https://doi.org/10.1145/3357384.3357895}
\showDOI{\tempurl}


\bibitem[Sun et~al\mbox{.}(2019a)]%
        {sun2019rotate}
\bibfield{author}{\bibinfo{person}{Zhiqing Sun}, \bibinfo{person}{Zhi-Hong
  Deng}, \bibinfo{person}{Jian-Yun Nie}, {and} \bibinfo{person}{Jian Tang}.}
  \bibinfo{year}{2019}\natexlab{a}.
\newblock \showarticletitle{Rotate: Knowledge graph embedding by relational
  rotation in complex space}.
\newblock \bibinfo{journal}{\emph{arXiv preprint arXiv:1902.10197}}
  (\bibinfo{year}{2019}).
\newblock


\bibitem[Tang and Wang(2018a)]%
        {tang2018personalized}
\bibfield{author}{\bibinfo{person}{Jiaxi Tang} {and} \bibinfo{person}{Ke
  Wang}.} \bibinfo{year}{2018}\natexlab{a}.
\newblock \showarticletitle{Personalized top-n sequential recommendation via
  convolutional sequence embedding}. In \bibinfo{booktitle}{\emph{Proceedings
  of the eleventh ACM international conference on web search and data mining}}.
  \bibinfo{pages}{565--573}.
\newblock


\bibitem[Tang and Wang(2018b)]%
        {tang2018ranking}
\bibfield{author}{\bibinfo{person}{Jiaxi Tang} {and} \bibinfo{person}{Ke
  Wang}.} \bibinfo{year}{2018}\natexlab{b}.
\newblock \showarticletitle{Ranking distillation: Learning compact ranking
  models with high performance for recommender system}. In
  \bibinfo{booktitle}{\emph{Proceedings of the 24th ACM SIGKDD international
  conference on knowledge discovery \& data mining}}.
  \bibinfo{pages}{2289--2298}.
\newblock


\bibitem[Wang and Blei(2011)]%
        {wang2011collaborative}
\bibfield{author}{\bibinfo{person}{Chong Wang} {and} \bibinfo{person}{David~M
  Blei}.} \bibinfo{year}{2011}\natexlab{}.
\newblock \showarticletitle{Collaborative topic modeling for recommending
  scientific articles}. In \bibinfo{booktitle}{\emph{Proceedings of the 17th
  ACM SIGKDD international conference on Knowledge discovery and data mining}}.
  \bibinfo{pages}{448--456}.
\newblock


\bibitem[Wang et~al\mbox{.}(2011)]%
        {wang2011cascade}
\bibfield{author}{\bibinfo{person}{Lidan Wang}, \bibinfo{person}{Jimmy Lin},
  {and} \bibinfo{person}{Donald Metzler}.} \bibinfo{year}{2011}\natexlab{}.
\newblock \showarticletitle{A cascade ranking model for efficient ranked
  retrieval}. In \bibinfo{booktitle}{\emph{Proceedings of the 34th
  international ACM SIGIR conference on Research and development in Information
  Retrieval}}. \bibinfo{pages}{105--114}.
\newblock


\bibitem[Weimer et~al\mbox{.}(2007)]%
        {weimer2007cofi}
\bibfield{author}{\bibinfo{person}{Markus Weimer}, \bibinfo{person}{Alexandros
  Karatzoglou}, \bibinfo{person}{Quoc Le}, {and} \bibinfo{person}{Alex Smola}.}
  \bibinfo{year}{2007}\natexlab{}.
\newblock \showarticletitle{Cofi rank-maximum margin matrix factorization for
  collaborative ranking}.
\newblock \bibinfo{journal}{\emph{Advances in neural information processing
  systems}}  \bibinfo{volume}{20} (\bibinfo{year}{2007}).
\newblock


\bibitem[Weston et~al\mbox{.}(2010)]%
        {weston2010large}
\bibfield{author}{\bibinfo{person}{Jason Weston}, \bibinfo{person}{Samy
  Bengio}, {and} \bibinfo{person}{Nicolas Usunier}.}
  \bibinfo{year}{2010}\natexlab{}.
\newblock \showarticletitle{Large scale image annotation: learning to rank with
  joint word-image embeddings}.
\newblock \bibinfo{journal}{\emph{Machine learning}} \bibinfo{volume}{81},
  \bibinfo{number}{1} (\bibinfo{year}{2010}), \bibinfo{pages}{21--35}.
\newblock


\bibitem[Wu et~al\mbox{.}(2021)]%
        {wu2021linear}
\bibfield{author}{\bibinfo{person}{Yongji Wu}, \bibinfo{person}{Defu Lian},
  \bibinfo{person}{Neil~Zhenqiang Gong}, \bibinfo{person}{Lu Yin},
  \bibinfo{person}{Mingyang Yin}, \bibinfo{person}{Jingren Zhou}, {and}
  \bibinfo{person}{Hongxia Yang}.} \bibinfo{year}{2021}\natexlab{}.
\newblock \showarticletitle{Linear-Time Self Attention with Codeword Histogram
  for Efficient Recommendation}. In \bibinfo{booktitle}{\emph{Proceedings of
  the Web Conference 2021}} (Ljubljana, Slovenia) \emph{(\bibinfo{series}{WWW
  '21})}. \bibinfo{publisher}{Association for Computing Machinery},
  \bibinfo{address}{New York, NY, USA}, \bibinfo{pages}{1262–1273}.
\newblock
\showISBNx{9781450383127}
\urldef\tempurl%
\url{https://doi.org/10.1145/3442381.3449946}
\showDOI{\tempurl}


\bibitem[Xu et~al\mbox{.}(2013)]%
        {xu2013cost}
\bibfield{author}{\bibinfo{person}{Zhixiang Xu}, \bibinfo{person}{Matt Kusner},
  \bibinfo{person}{Kilian Weinberger}, {and} \bibinfo{person}{Minmin Chen}.}
  \bibinfo{year}{2013}\natexlab{}.
\newblock \showarticletitle{Cost-sensitive tree of classifiers}. In
  \bibinfo{booktitle}{\emph{International conference on machine learning}}.
  PMLR, \bibinfo{pages}{133--141}.
\newblock


\bibitem[Xu et~al\mbox{.}(2014)]%
        {xu2014classifier}
\bibfield{author}{\bibinfo{person}{Zhixiang Xu}, \bibinfo{person}{Matt~J
  Kusner}, \bibinfo{person}{Kilian~Q Weinberger}, \bibinfo{person}{Minmin
  Chen}, {and} \bibinfo{person}{Olivier Chapelle}.}
  \bibinfo{year}{2014}\natexlab{}.
\newblock \showarticletitle{Classifier cascades and trees for minimizing
  feature evaluation cost}.
\newblock \bibinfo{journal}{\emph{The Journal of Machine Learning Research}}
  \bibinfo{volume}{15}, \bibinfo{number}{1} (\bibinfo{year}{2014}),
  \bibinfo{pages}{2113--2144}.
\newblock


\bibitem[Ying et~al\mbox{.}(2018)]%
        {ying2018graph}
\bibfield{author}{\bibinfo{person}{Rex Ying}, \bibinfo{person}{Ruining He},
  \bibinfo{person}{Kaifeng Chen}, \bibinfo{person}{Pong Eksombatchai},
  \bibinfo{person}{William~L Hamilton}, {and} \bibinfo{person}{Jure Leskovec}.}
  \bibinfo{year}{2018}\natexlab{}.
\newblock \showarticletitle{Graph convolutional neural networks for web-scale
  recommender systems}. In \bibinfo{booktitle}{\emph{Proceedings of the 24th
  ACM SIGKDD international conference on knowledge discovery \& data mining}}.
  \bibinfo{pages}{974--983}.
\newblock


\bibitem[Zhang et~al\mbox{.}(2013)]%
        {zhang2013optimizing}
\bibfield{author}{\bibinfo{person}{Weinan Zhang}, \bibinfo{person}{Tianqi
  Chen}, \bibinfo{person}{Jun Wang}, {and} \bibinfo{person}{Yong Yu}.}
  \bibinfo{year}{2013}\natexlab{}.
\newblock \showarticletitle{Optimizing top-n collaborative filtering via
  dynamic negative item sampling}. In \bibinfo{booktitle}{\emph{Proceedings of
  the 36th international ACM SIGIR conference on Research and development in
  information retrieval}}. \bibinfo{pages}{785--788}.
\newblock


\bibitem[Zhou et~al\mbox{.}(2019)]%
        {zhou2019deep}
\bibfield{author}{\bibinfo{person}{Guorui Zhou}, \bibinfo{person}{Na Mou},
  \bibinfo{person}{Ying Fan}, \bibinfo{person}{Qi Pi}, \bibinfo{person}{Weijie
  Bian}, \bibinfo{person}{Chang Zhou}, \bibinfo{person}{Xiaoqiang Zhu}, {and}
  \bibinfo{person}{Kun Gai}.} \bibinfo{year}{2019}\natexlab{}.
\newblock \showarticletitle{Deep interest evolution network for click-through
  rate prediction}. In \bibinfo{booktitle}{\emph{Proceedings of the AAAI
  conference on artificial intelligence}}, Vol.~\bibinfo{volume}{33}.
  \bibinfo{pages}{5941--5948}.
\newblock


\bibitem[Zhou et~al\mbox{.}(2018)]%
        {zhou2018deep}
\bibfield{author}{\bibinfo{person}{Guorui Zhou}, \bibinfo{person}{Xiaoqiang
  Zhu}, \bibinfo{person}{Chenru Song}, \bibinfo{person}{Ying Fan},
  \bibinfo{person}{Han Zhu}, \bibinfo{person}{Xiao Ma},
  \bibinfo{person}{Yanghui Yan}, \bibinfo{person}{Junqi Jin},
  \bibinfo{person}{Han Li}, {and} \bibinfo{person}{Kun Gai}.}
  \bibinfo{year}{2018}\natexlab{}.
\newblock \showarticletitle{Deep interest network for click-through rate
  prediction}. In \bibinfo{booktitle}{\emph{Proceedings of the 24th ACM SIGKDD
  international conference on knowledge discovery \& data mining}}.
  \bibinfo{pages}{1059--1068}.
\newblock


\bibitem[Zhu et~al\mbox{.}(2018)]%
        {zhu2018learning}
\bibfield{author}{\bibinfo{person}{Han Zhu}, \bibinfo{person}{Xiang Li},
  \bibinfo{person}{Pengye Zhang}, \bibinfo{person}{Guozheng Li},
  \bibinfo{person}{Jie He}, \bibinfo{person}{Han Li}, {and}
  \bibinfo{person}{Kun Gai}.} \bibinfo{year}{2018}\natexlab{}.
\newblock \showarticletitle{Learning tree-based deep model for recommender
  systems}. In \bibinfo{booktitle}{\emph{Proceedings of the 24th ACM SIGKDD
  International Conference on Knowledge Discovery \& Data Mining}}.
  \bibinfo{pages}{1079--1088}.
\newblock


\end{thebibliography}

\newpage
\appendix
\section{Proofs of Theoretical Results}\label{sec:appd_proof}

\subsection{Proof of Theorem \ref{thm:kl_convergence}}
\begin{proof}
	\begin{displaymath}
			\begin{aligned}
			&D_{KL}\Big(P_\phi(\cdot|c)\parallel P_\theta(\cdot|c)\Big)\\
			 =& \sum_{i\in \mathcal{I}} P_\phi(i|c) \log \frac{P_\phi(i|c)}{P_\theta(i|c)}\\
			=&\mathbb E_{i\sim P_\phi(\cdot|c)}[R_\phi(i,c)-M_\theta(i,c)] + \log Z_\theta - \log Z_\phi\\
			=&\mathbb E_{i\sim P_\phi(\cdot|c)}[\Delta_i^c] - \log \frac{\sum_{j\in \mathcal{I}}\exp(R_\phi(j,c))}{\sum_{j\in \mathcal{I}}\exp(M_\theta(j,c))}\\
			=&\mathbb E_{i\sim P_\phi(\cdot|c)}[\Delta_i^c] - \log \sum_{j\in \mathcal{I}}P_\theta(j|c)\frac{\exp(R_\phi(j,c))}{\exp(M_\theta(j,c))}\\
			=&\mathbb E_{i\sim P_\phi(\cdot|c)}[\Delta_i^c]-\log \mathbb E_{i\sim P_\theta(\cdot|c)}[\exp(\Delta_i^c)]\\
			\underset{L\rightarrow \infty}{\overset{a.s.}{\leftarrow}}& \sum_{k=1}^n {P}_\phi^\mathcal{S}(o_k|c) \Delta_{o_k}^c - \log \sum_{k=1}^n {P}_\theta^\mathcal{S}(o_k|c) \exp{(\Delta_{o_k}^c)} \\
			=&\mathbb E_{i\sim {P}_\phi^{\mathcal{S}}(\cdot|c)}[\Delta_i^c]-\log \mathbb E_{i\sim {P}_\theta^\mathcal{S}(\cdot|c)}[\exp(\Delta_i^c)]\\
		\end{aligned}
	\end{displaymath}
where $\Delta_i^c\triangleq R_\phi(i,c)-M_\theta(i,c)$. ${P}_\phi^\mathcal{S}(j|c)=\frac{\exp(\tilde{R}_\phi(j,c))}{\sum_{i\in \mathcal{S}}\exp(\tilde{R}_\phi(i,c))}$ and ${P}_\theta^\mathcal{S}(j|c)=\frac{\exp(\tilde{M}_\theta(j,c))}{\sum_{i\in \mathcal{S}}\exp(\tilde{M}_\theta(i,c))}$, where $\tilde{R}_\phi(i,c) =R_\phi(i,c) - \log \tilde{Q}(i|c)$ and $\tilde{M}_\theta(i,c) = M_\theta(i,c) - \log \tilde{Q}(i|c)$. The almost sure convergence is based on the results in self-normalized importance sampling, which can be proved by simply applying the strong law of large number.
\begin{displaymath}
\begin{aligned}
	&D_{KL}\Big({P}_\phi^\mathcal{S}(\cdot|c)\parallel {P}_\theta^\mathcal{S}(\cdot|c)\Big)\\
	=& \sum_{i\in \mathcal{S}} {P}_\phi^\mathcal{S}(i|c)\log \frac{{P}_\phi^\mathcal{S}(i|c)}{{P}_\theta^\mathcal{S}(i|c)}\\
	=& \mathbb E_{i\sim {P}_\phi^\mathcal{S}(\cdot|c)}[\tilde{R}_\phi(i,c)-\tilde{M}_\theta(i,c)] +\log Z^\mathcal{S}_\theta - \log Z^\mathcal{S}_\phi\\
	=&\mathbb E_{i\sim {P}_\phi^\mathcal{S}(\cdot|c)}[R_\phi(i,c)-M_\theta(i,c)] - \log \frac{\sum_{j\in \mathcal{S}}\exp(\tilde{R}_\phi(i,c))}{\sum_{j\in \mathcal{S}}\exp(\tilde{M}_\theta(i,c))}\\
	=&\mathbb E_{i\sim {P}_\phi^\mathcal{S}(\cdot|c)}[\Delta_i^c]-\log \sum_{j\in \mathcal{S}}{P}_\theta^\mathcal{S}(j|c)\frac{\exp(\tilde{R}_\phi(j,c))}{\exp(\tilde{M}_\theta(j,c))} \\
	=&\mathbb E_{i\sim {P}_\phi^\mathcal{S}(\cdot|c)}[\Delta_i^c]-\log \mathbb E_{i\sim {P}_\theta^\mathcal{S}(\cdot|c)}[\exp(\tilde{R}_\phi(i,c)-\tilde{M}_\theta(i,c))]\\
	=&\mathbb E_{i\sim {P}_\phi^\mathcal{S}(\cdot|c)}[\Delta_i^c]-\log \mathbb E_{i\sim {P}_\theta^\mathcal{S}(\cdot|c)}[\exp(\Delta_i^c)]\\
\end{aligned}
\end{displaymath}
Therefore, $D_{KL}\Big({P}_\phi^\mathcal{S}(\cdot|c)\parallel {P}_\theta^\mathcal{S}(\cdot|c)\Big)$ is an asymptotically unbiased estimation of $D_{KL}\Big(P_\phi(\cdot|c)\parallel P_\theta(\cdot|c)\Big)$.
\end{proof}
\subsection{Proof of Corollary \ref{thm:kl_corollary}}
\begin{proof}
\begin{displaymath}
	\begin{aligned}
		&D_{KL}(P_\phi\parallel P_\theta)\\
		=& \sum_{i\in \mathcal{I}} P_\phi(i|c) \log \frac{P_\phi(i|c)}{P_\theta(i|c)}\\
		=&\mathbb E_{i\sim P_\phi(\cdot|c)}[\Delta_i^c]-\log \mathbb E_{i\sim P_\theta(\cdot|c)}[\exp(\Delta_i^c)]\\
		\approx&  \frac{\sum_{i\in \mathcal{S}}\exp(R_\phi(i,c) - \log Q(i|c))\Delta_{o_k}^c}{\sum_{i\in \mathcal{S}}\exp(R_\phi(i;c) - \log Q(i|c))}  - \log \sum_{i\in \mathcal{S}} \exp{(\Delta_{i}^c)} + \log L \\
		=&\frac{\sum_{i\in \mathcal{S}}\exp(R_\phi(i;c) -M_\theta(i,c))\Delta_{i}^c}{\sum_{i\in \mathcal{S}}\exp(R_\phi(i,c) - M_\theta(i,c))}  - \log \sum_{i\in \mathcal{S}} \exp{(\Delta_{i}^c)} + \log L \\
		=&\frac{\sum_{i\in \mathcal{S}}\exp(\Delta_{i}^c)\Delta_{i}^c}{\sum_{i\in \mathcal{S}}\exp(\Delta_{i}^c)}  - \log \sum_{i\in \mathcal{S}} \exp{(\Delta_{i}^c)} + \log L \\
		=&\sum_{j\in \mathcal{S}}\frac{\exp(\Delta_{j}^c)}{\sum_{i\in \mathcal{S}}\exp(\Delta_{i}^c)} \log\frac{\exp \Delta_{j}^c}{\sum_{i\in \mathcal{S}} \exp{(\Delta_{i}^c)}}  + \log L \\
		=& \log L - H({\normalfont \text{softmax}}(\boldsymbol{\Delta}^c))
	\end{aligned}
\end{displaymath}
\end{proof}

\section{Experiments}

\subsection{Baseline Methods}\label{sec:appd_baseline}
Following baseline approaches are retriever methods in general recommendation(the first four) and sequential recommendation(the last four).
\begin{itemize}[leftmargin=*]
    \item \textbf{BPR~\cite{rendle2009bpr}:} BPR is a matrix factorization method with a loss function based on Bayes equation.
    \item \textbf{NCF~\cite{he2017neural}:} NCF combines matrix factorization method and MLP to get deeper interaction in score. We don't use pretraining technique in our experiments.
    \item \textbf{LogisticMF~\cite{Johnson2014LogisticMF}:} LogisticMF is a probabilistic model for matrix factorization, whose optimization goal is to increase the probability for interacted items and to decrease the probability for uninteracted items. 
    \item \textbf{DSSM~\cite{huang2013learning}:} DSSM is a two-tower model designed for information retrieval, which models query and key by simple MLP layer and then gets socres by inner product operation. Here we treat user and item as query and key respectively.
    \item \textbf{GRU4Rec:~\cite{hidasi2015session}} GRU4Rec uses a one-layer GRU to obtain latent vector of user's behavior sequence.
    \item \textbf{Caser:~\cite{tang2018personalized}} Caser is a famous model which apply several CNN units with kernels of different size to capture user intent.   
    \item \textbf{BERT4Rec:~\cite{sun2019bert4rec}} BERT4Rec applies the cloze task in sequential recommendation, which is firstly proposed in BERT in language modeling. And it captures user intent with a Transformer encoder.
    \item \textbf{SASRec~\cite{kang2018self}:} SASRec models user's behavior with Transformer encoder, where multi-head attention mechanism is attached to great importance.
\end{itemize}
Following baseline approaches are cooperative methods compared with CoRR.
\begin{itemize}[leftmargin=*]
    \item \textbf{Independent:} Two-Stage is a simple combination of SASRec (as retriever) and DIN (as ranker), where the two models are trained independently. We use a two-step strategy to predict: retrieve items by SASRec and then rank those items by DIN. Other variants of the Two-Stage based approaches, such as cascading structures, can be attached in Section~\ref{sec:help_ranker} and~\ref{sec:help_retriever}.
    \item \textbf{ICC~\cite{gallagher2019joint}:} ICC is a joint training method of cascade ranking. The score of a pair of query and key is actually the weighted sum of scores in each stage. And the weight of higher(more closed to retrieval) stage is larger. 
    \item \textbf{RankFlow~\cite{qin2022rankflow}:} RankFlow is a recently proposed joint training method, which consists of two alternative training flow(self learning and tutor learning). In self learning flow, upstream models provide top-k items as negatigves for the downstream models. And the downstream models distill weak signals to upstream models in the tutor learning flow.
\end{itemize}

\subsection{Implementation Details}\label{sec:implement_appd}

In general recommendation, the hidden size for MLP layer in DeepFM is set as [128,128,128]. In sequential recommendation, 
as for SASRec, we use one Transformer encoder layer with two-head attention blocks. The feedforward dimension is set as 128. As for DIN, We adopt Sigmoid as the activation function and set the hidden size as 64 for activation unit. And the final MLP prediction layer size is set as [200, 80].  The dimension of embedding is set as 64 in all models. The learning rate is chosen from $\{0.0001, 0.001, 0.01\}$ and the weight decay ${l}_2$ is chosen from $\{0,10^{-6},10^{-5}, 10^{-4}\}$. The dropout output rate is set as $0.3$. The batch size is set to 512 in the Gowalla, Amazon and MovieLens datasets, and 2048, 1024 in Taobao for general recommendation and sequential recommendation respectively. During the training process, 20 items are sampled as negative samples with dynamic hard negative sampler MIDX\_Uni\cite{chen2021fast} for training retriever and ranker.

% \subsection{Student's T-test}\label{sec:appd_ttest}
% In order to demonstrate CoRR is indeed better than baseline methods, we apply Student's t-test to check whether CoRR is better than other methods in all metrics and datasets both in general recommendation and sequential recommendation. The p-values are shown in Table \ref{tab:p-value}:

% The p-values in the table are almost smaller than $10^{-3}$, which means the assumption that CoRR outperforms other methods holds with high probability. Therefore, we conclude that CoRR is significantly better than retrievers and other jointly methods in all datasets. 

\begin{table}[h]
    \caption{Comparison among various knowledge distillation loss. ($\times 10^{-2}$)}
    %\vspace{-0.3cm}
    % \setlength\tabcolsep{3pt}
    \begin{tabular}{l|c|c|c|c}
    \toprule
        Distill Loss & w/o Distill & RankDistill &  CoDistill & Ours \\ \midrule
        Recall@10 & 13.42 & 13.66 & 13.49 & 14.46 \\ \midrule
        NDCG@10 & 6.97 & 7.88 & 7.83 & 8.31 \\ \midrule
        MRR@10 & 6.00 & 6.17 & 6.11 & 6.44 \\ \bottomrule
    \end{tabular}
    \label{tab:kl_loss2}
\end{table}

\begin{figure}[ht]
    \centering
    \subfigure[Amazon-RankFlow]{
        \includegraphics[width=0.48\columnwidth]{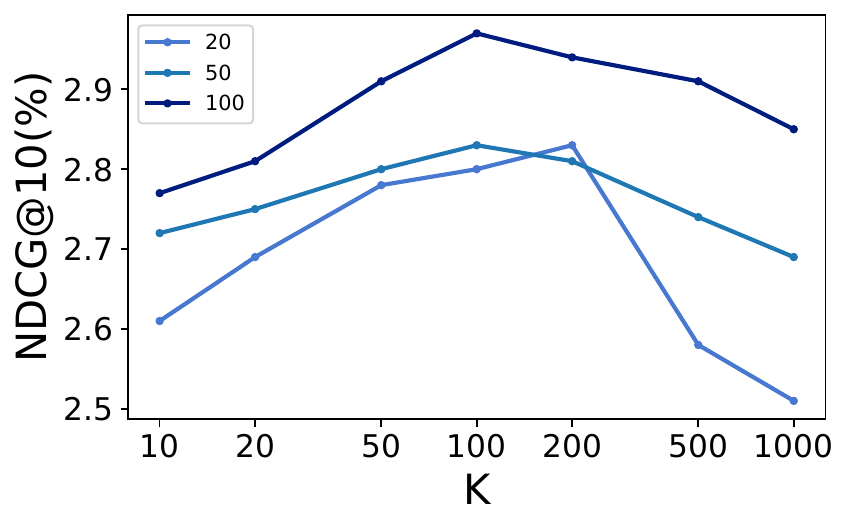}
    }
    \hspace{-0.3cm}
    \vspace{-8pt}
    \subfigure[Amazon-CoRR]{
        \includegraphics[width=0.48\columnwidth]{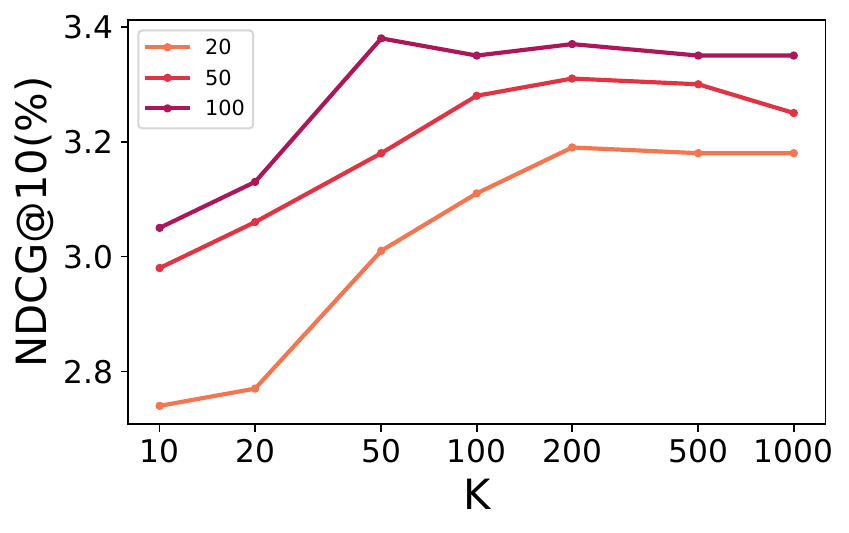}
    }
    \vspace{-8pt}
    \subfigure[Gowalla-RankFlow]{
        \includegraphics[width=0.48\columnwidth]{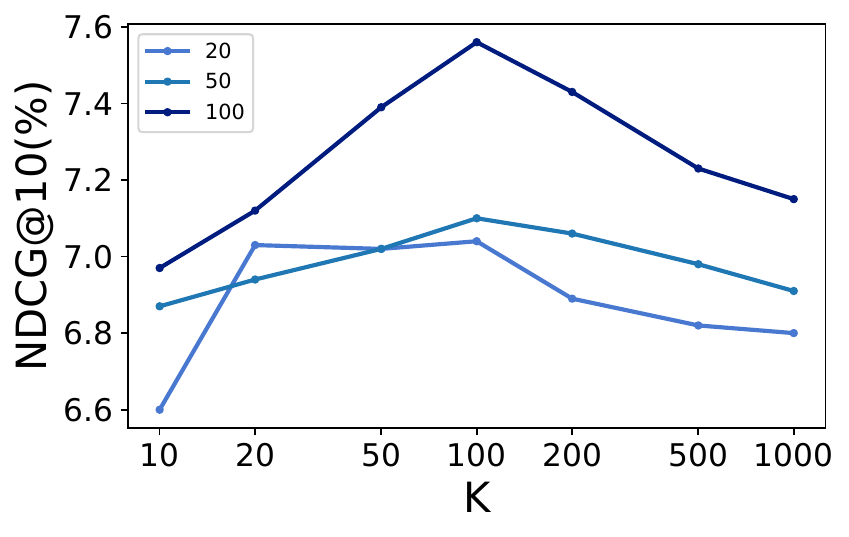}
    }
    \hspace{-0.3cm}
    \vspace{-8pt}
    \subfigure[Gowalla-CoRR]{
        \includegraphics[width=0.48\columnwidth]{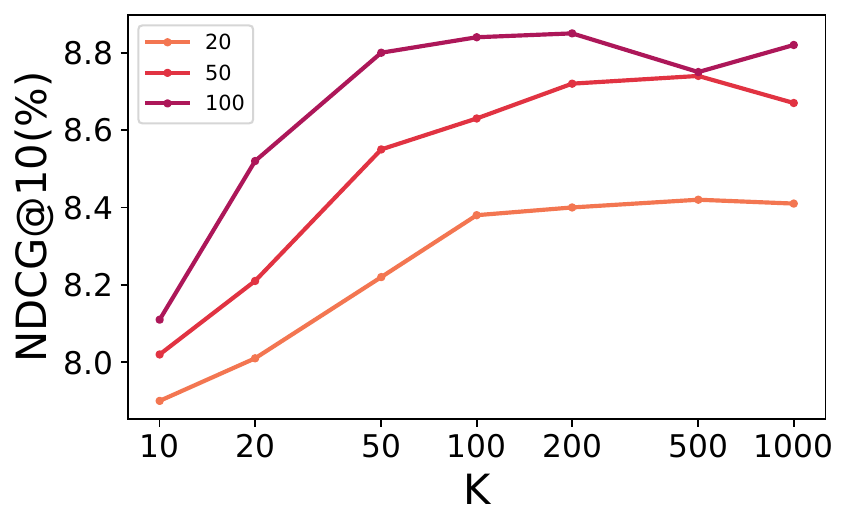}
    }
    \caption{Sensitivity w.r.t. Retrieved Items Number}
    \Description[Sensitivity w.r.t. Retrieved Items Number]{Fully described in the Appendix B.4.}
    \label{fig:topk}
\end{figure}

\subsection{Comparison of Different Distillation Loss}
According to results in Section~\ref{sec:help_retriever}, knowledge distillation indeed plays an important role in improving the retriever and the whole framework. Ranking-based distillation loss is popular in knowledge distillation in recommendation, which usually attaches larger weights to the items with higher rank. To further study the effectiveness of our distillation loss, we compare several ranking-based distillation strategies with ours. They include:
\begin{itemize}
    \item RankDistill~\cite{tang2018ranking}: ranks of topk items are used as the weights for negative log likelihood loss.
    \item CoDistill~\cite{lee2019collaborative}: items are sampled by a ranking-based sampling strategy and then a cross entropy loss is used for closing the scores of two models.
\end{itemize}
 The results shown in Table~\ref{tab:kl_loss2} demonstrate that: First, the addition of any one of the three distillation losses is able to enhance the performance of the whole framework. CoDistill loss, whose improvement is smaller, even outperforms w/o Distill with a relative 12.34\% improvement on NDCG@10. Besides, our sampled KL loss obviously outperforms the other two losses. Our KL loss achieves 5.86\% and 7.19\% improvements on Recall@10 compared with RankDistill and CoDistill respectively, indicating the superiority of the asymptotic sampled kl-divergence.

\subsection{Sensitivity of the Retrieval Cutoff}\label{sec:topk_appd}
As mentioned in Section~\ref{sec:details}, a two-stage strategy is applied in prediction: K candidates are retrieved by retriever firstly and then refined by the ranker to get the final recommendation results. The performance may be affected by the retrieval cutoff due to item distribution shift between training and inference of the ranker. To further verify the sensitivity to the retrieval cutoff, we conduct experiments on RankFlow and CoRR by varying the cutoff K in the inference stage within \{10, 20, 50, 100, 200, 500, 1000\} and varying the number of negatives in \{20, 50, 100\}. The results are reported in Figure~\ref{fig:topk}. 
% Since two-stage str ategy is applied in recommendation as mentioned in Section~\ref{sec:details}, the number of retrieved candidates 

These figures illustrate that the performance gets better as K increases from 10 to 100 for CoRR and RankFlow, which is explained by the limited expressiveness of the retriever. However, the tendencies vary between RankFlow and CoRR when K becomes larger (from 100 to 1000). The performance of CoRR almost holds steady as K increases while RankFlow suffers a severe drop since the rankers face highly shifted item distributions from the training stage. This indicates that CoRR is capable of addressing item distribution shift.

\begin{figure}[ht]
    \centering
    \subfigure[Cluster Number]{
        \includegraphics[width=0.40\columnwidth]{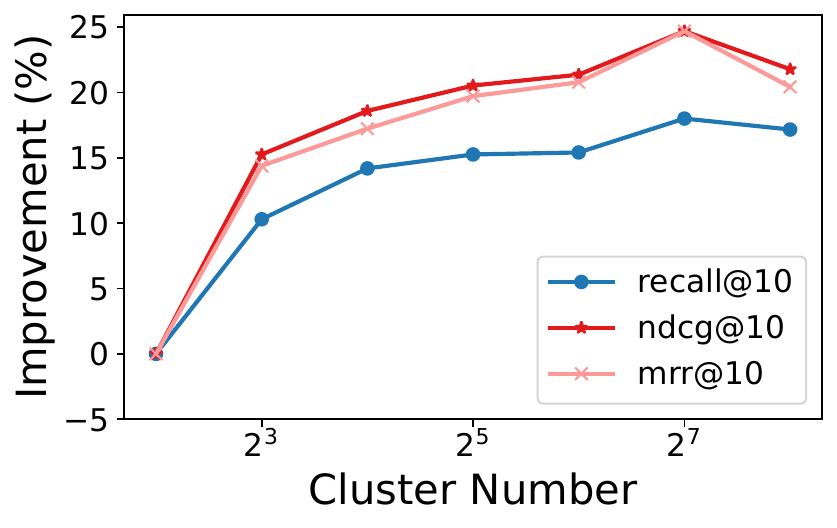}
    }
    \hspace{-0.3cm}
    \vspace{-8pt}
    \subfigure[Temperature]{
        \includegraphics[width=0.50\columnwidth]{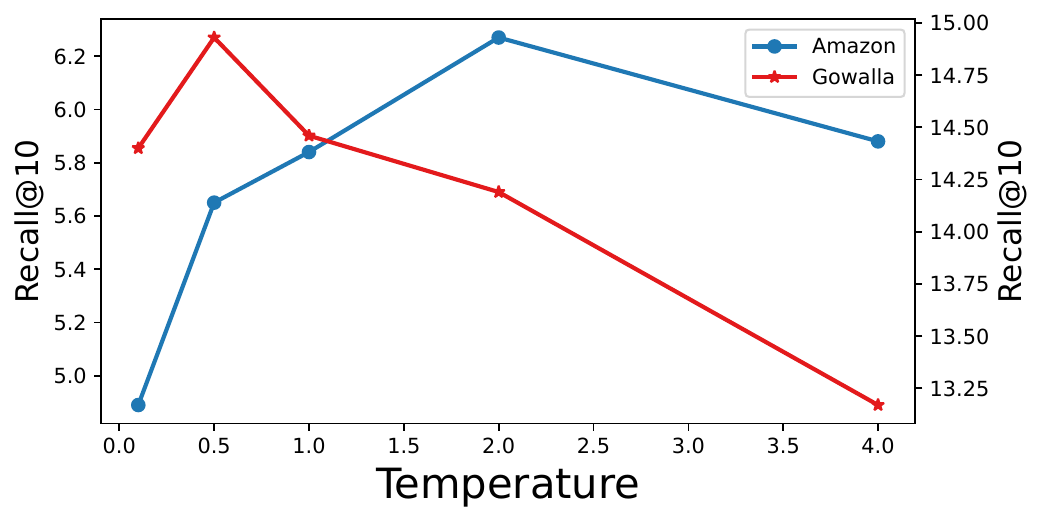}
    }
    \caption{Sensitivity w.r.t. Cluster Number and Temperature}
    \label{fig:sens_others}
    \Description[Sensitivity w.r.t. Cluster Number and Temperature]{Fully described in the Appendix B.5\-B.6.}
\end{figure}

\subsection{Sensitivity of the Cluster Number}
As mentioned in Section~\ref{sec:resample}, our scalable and adaptive sampling strategy adopts clustering technique to build index. As the key parameter of clustering, the cluster number (K) would be related to the effectiveness of the sampling. Therefore, we conduct experiments by varing cluster number within \{4,8,16,32,64,128,256\} on Gowalla. The results are shown in Figure~\ref{fig:sens_others}(a).

The figure illustrates that as the number of clusters increases, the ranking performance first improves and then saturates. When the cluster number is too small(K=4), there would be amounts of items in the same cluster, which would result in more information loss. When the cluster number is large enough, the performance is relatively insensitive to K.

\subsection{Sensitivity of the Temperature}
As discussed in Section~\ref{sec:resample}, temperature T controls the balance between hardness and randomness of negative samples. We conduct experiments on CoRR by varying the temperature T within \{0.1,0.5,1,2,4\}. The results are reported in Figure~\ref{fig:sens_others}(b).

When T increases, the sampling distribution is more closed to uniform distribution and the sampler concerns more about randomness. On the contrary, the sampler concerns more about hardness. The results on Amazon and Gowalla dataset show that T=1 is a good option for the final performance.

\end{document}